\documentclass[12pt,prb,aps,preprint]{revtex4}
\usepackage{graphicx,epsfig}
\usepackage{floatflt}
\usepackage{enumerate}

\begin{document}

\title{Multiple-choice test of energy and momentum concepts}
\author{Chandralekha Singh}
\author{David Rosengrant}
\affiliation{Department of Physics and Astronomy, University of Pittsburgh, Pittsburgh, PA 15260}

\begin{abstract}
We investigate student understanding of energy and momentum
concepts at the level of introductory physics by designing and
administering a 25-item multiple choice test and conducting
individual interviews. We find that most students have difficulty
in qualitatively interpreting basic principles related to energy
and momentum and in applying them in physical situations. 
\end{abstract}

\maketitle

\section{Introduction}
Research-based multiple-choice tests can be powerful probes of
student understanding. They are easy and economical to administer
and to grade, have objective scoring, and are amenable to
statistical analysis that can be used to compare student
populations or instructional methods. A major drawback is that the
thought processes are not revealed completely by the answers
alone. However, when combined with student interviews,
well-designed tests are powerful tools for educational assessment.
Conceptual multiple-choice tests have already been designed to
assess student understanding of force, kinematics graphs, heat,
circuits, and electricity and magnetism.\cite{mdt} They suggest
that students' knowledge of physics often is fragmented and
context-dependent, and that students have many common difficulties.

Energy and momentum are two of the most fundamental concepts in
physics.\cite{alan} In the lower-level physics courses, they are
typically introduced after Newton's laws. We have designed and
administered a research-based 25-item multiple-choice test and
conducted individual interviews to explore student understanding
of energy and momentum concepts.\cite{singh} We are interested in
understanding the difficulties students have in interpreting these
concepts and in correctly identifying and applying them in
different physical situations. We also wish to know the extent to
which the difficulties are universal, and if there is a
correlation with student preparation (for example, calculus or
algebra background). The identification of student difficulties
can help in designing instructional tools that address them. Part
of the rationale for combining energy and momentum concepts is to
investigate the extent to which students can identify the relevant
concept in a particular situation.

\section{Test Design}
During the test design, we paid particular attention to the
important issues of reliability and validity.\cite{nitko}
Reliability refers to the relative degree of consistency in scores
between testing if the test procedures are repeated for an
individual or group. Validity refers to the appropriateness of the
test score interpretation. The test design began with the
development of a test blueprint that provided a framework for
planning decisions about the desired test attributes. We tabulated
the scope and extent of the content covered and the level of
cognitive complexity desired. The specificity of the test
blueprint was useful for creating questions. The energy concepts included the work-energy
theorem, conservation of mechanical energy, and work done by
gravitational and frictional forces. The momentum concepts
included the definition of momentum, impulse-momentum theorem
(impulse was defined explicitly in the test), and conservation of
momentum with examples from elastic and inelastic collisions. We
also planned to evaluate student understanding of the concept of
``system" in various contexts. We also planned to include one question that
explicitly required the application of both energy and momentum
concepts. Energy and momentum questions
pertaining to simple harmonic motion, explicit mention of
conservative and non-conservative forces and center-of-mass
reference frame were deliberately excluded, because we wanted the
test content to be commensurate with the curriculum in most
calculus- and algebra-based courses for science and engineering
majors. We classified the cognitive complexity using a simplified
version of Bloom's taxonomy:\cite{bloom} specification of
knowledge, interpretation of knowledge and drawing inferences, and
applying knowledge to different situations. Then, we outlined
 a description of
conditions/contexts within which the various concepts would be tested and a criteria for good performance in each case
. The tables
of content and cognitive complexity along with the criteria for good performance were shown to five physics
faculty members at the University of Pittsburgh for review.
Modifications were made to the weights assigned to various
concepts and to the performance criteria based upon the feedback
from the faculty about their appropriateness. The performance
criteria was used to convert the description of conditions/contexts within which the concepts would be tested to make approximately 50 free-response questions. These questions
required students to provide their reasoning for their responses. 

The free-response questions were administered (in groups of 10 or 20)
to students in the calculus- and algebra-based courses at the
University of Pittsburgh in the Fall of 1998. Often, some
students in a class were given one set of questions and others were
given another set in order to sample student responses on most of
the questions. We also tape-recorded interviews with 10 introductory
student volunteers using the think-aloud protocol.\cite{chi} Forty
multiple-choice items (questions) were then designed using the most frequent
student responses for the free-response questions and interviews as a
guide for making the alternative distractor choices. Four alternative choices have typically been found to be optimal~\cite{nitko} and we chose the
four distractors to conform to the common difficulties to increase the discriminating properties of the
items. Ten physics faculty members
and postdocs were asked to review the multiple-choice items
and comment on their appropriateness and relevance for
introductory physics and to detect ambiguity in item wording. An
item review form was developed to aid the faculty in reviewing the
items. The faculty also classified each item on a scale from very
appropriate to least appropriate. Further modifications were made
based upon their recommendation. Then, a multiple-choice test was
assembled using 28 items which closely matched the initial table
delineating the scope of the content and cognitive complexity. The
same faculty members who reviewed the items were shown the test
and some modifications were made.

The test was administered as a 50-minute post-test (after
instruction) to several hundred students in calculus- and
algebra-based courses at the University of Pittsburgh in Spring
1999. Seven student volunteers who had taken the test were
interviewed using the think-aloud protocol. Item analysis of
student responses was performed to judge the quality of each item.
In addition to the calculation of item difficulty and
discrimination,\cite{nitko} the item analysis included creating a
table to count the number of students selecting each distractor in
the upper and lower quartiles. This analysis was very useful to
determine whether individual items and distractors functioned as
expected. Based upon the item analysis and interviews, the test
items were modified further. The number of items in the test was
decreased to 25. Including the spring and fall 1999, 2000 and 2001
the test has been administered to more than 3000 students from
approximately 30 calculus- and algebra-based courses in different
colleges and universities. Some classes administered the test both
as a pre-/post-test to assess students' conceptions of energy and
momentum before instruction, the effectiveness of the
instruction, and the effect of pre-test on post-test. Every
semester an item analysis was performed after its administration,
a few students were interviewed, and some items were slightly
modified. We have conducted a total of 34 one-hour interviews
during the test development process. In Fall 2000, the test was
administered to graduate students enrolled in a teaching methods
class. Their average score was greater than 80\% and the
reliability coefficient\cite{nitko} $\alpha$ (which is a measure
of the internal consistency of the test\cite{nitko}) was greater
than 0.8. 

\section{Results}
The final version of the test is included in Appendix~\ref{appa}. The test was administered before the instruction on
energy and momentum concepts (pre-test) only in some classes. The
average pre-test scores typically ranged from 24\%--36\% depending
upon the class type, for example, calculus versus algebra (higher
for calculus), and whether the test was administered before or
after any instruction on linear kinematics/dynamics (some questions
can be answered without explicitly invoking energy or momentum
concepts).

We first discuss the post-test results of the version administered
to 1356 students in Fall 2000. The reliability coefficient
$\alpha$ for the calculus-based classes was
slightly more than 0.75 (1170 students), while that for the
algebra-based classes was 0.68 (186 students). Figures 1 and 2
show the post-test statistics for the calculus-based courses (1170
students) including the honors (two) and active-engagement classes
(one). The average post-test score was 49\% (standard deviation $\sigma=$ 18\%).
Figure 1 shows a plot of the item difficulty (the fraction of
students who got the item correct), which ranged from 0.2 to 0.8.
This range is consistent with our goal because we want to explore
student performance on items with a variety of cognitive
complexity. Figure~2 shows a plot of the point biserial
discrimination (PBD)\cite{nitko} for individual items which
measures the ability of an item to distinguish between high (those
who performed well on the test overall) and low performing
students. The PBD ranged from 0.21 to 0.48 with only one item with
PBD less than 0.28, which is good by the standards of achievement
test design.\cite{nitko} [xx it looks like you left out some
references to Table I and II which were in the hard copy from
Redish xx] The average score on the items focusing mainly on the
energy concepts (14 total) was 46\% ($\sigma=$ 20\%), while that on
items focusing mainly on momentum concepts (10 items) was 55\%
($\sigma=$ 22\%). There was a reasonable correlation between the
performance on the energy questions and momentum questions (Pearson
correlation\cite{nitko} 0.54). We performed a factor analysis
(principal component method)\cite{factor} which identified six
factors with a variance greater than 4\%. However, the largest of
these factors accounted for less than 15\% of the variance. This
variance is very small for the first factor so none of the factors
are meaningful. Although the analysis of variance
(ANOVA)\cite{nitko} shows statistically significant differences
between several calculus-based classes at the level of
$\alpha=0.05$, the effect size
$d$ is small ($d<0.35$) for all pairs except those involving an
honors class or an active-engagement class.\cite{cohen} Also,
ANOVA shows a statistically significant difference between the
calculus-based (excluding the honors and active-engagement
classes) and algebra-based classes in terms of the total average score,
but again the effect size is small. ANOVA on individual items
shows that the differences between the calculus- and algebra-based
classes are not statistically significant for 12 of the 25 test
items. 

\subsection{Pre/Post Instruction Analysis}
To assess the effect of instruction, we now discuss results
(unmatched) from three calculus-based classes for which both
pre-/post-tests were administered. A total of 352 and 336 students
from the three classes took the pre-test and post-test in Fall
2001 and the average scores were 33.6\% and 51.7\% respectively.
The average post-test score was more than 50\% on 15 out of the 25
items. Table~I shows the average percent correct on the
pre-/post-test and the normalized gain
$g=(s_f-s_i)/(100-s_i)$, where $s_i$ and $s_f$ are the
pre-/post-test scores in percent)\cite{hake} on each test item
separately for students whose total test
scores were in the upper 25\% (group H), lower 25\% (group L), and
all students (A). The average pre-test scores of students in the
upper and lower quartiles were 49.8\% and 19.2\% and the average
post-test scores in these two groups were 70.6\% and 32.2\%
respectively. Thus, students in the upper quartile gained
significantly more than those in the lower quartile with an
average gain of 0.45 for group H, 0.16 for group L, and
0.29 for all students A. The upper and lower quartiles were
selected based the average scores on the pre and post tests.
Table~II shows the distribution of alternative choices for the
pre-/post-tests. It shows the types of preconceptions students had
before instruction on the energy and momentum concepts and how the
conceptions changed after the instruction. Comparison of the
pre-/post-tests in Table~II shows that instruction improved
student conceptions in some cases, but not in others.

The test results reveal that students lack a coherent understanding
of energy and momentum concepts and have difficulty applying them
to different physical situations. They often focused on the
surface features and were distracted by irrelevant
details.\cite{chi3} During the interview,
some students suggested that the lack of numerical values for the
parameters prevented them from checking if they are on the right
track. Many students did not realize that work and energy are
scalar quantities and momentum is a vector quantity. Questions
involving work-energy or impulse-momentum theorems were typically
perceived to be more difficult than those involving their special
cases: the conservation of mechanical energy and momentum
respectively. However, students had great difficulty in using the
conservation principles appropriately in many situations. In the
following, we discuss some insightful features of student responses
by loosely clustering items based upon the relevant physics
concepts. Although the same item can be categorized in
different ways, we choose to discuss them within a particular
cluster.

\medskip{\it Questions related to energy concepts}.
Items 2, 4, 15, and 22 can all be answered based upon the
conservation of mechanical energy, because there is no work done by
non-conservative forces. (The normal force is present in the
``slide" related items (2) and (15), but it is perpendicular to the
direction of motion of the object, so it does no work on it.)
Students performed very differently on these items and did not
perform any better on items (4) and (22) which involved only the
gravitational force compared to items 2 and 15 which had a normal
force present. Interviews show that in many physical situations
students knew ``what" but did not know ``why.'' For example, in
response to item 4, several interviewed students said that the
balls thrown from a cliff will reach the ground at the same speed,
regardless of the angle of projection, but they could not justify
their answers based either upon energy or kinematic
considerations. Item 15 elicits strong belief that ``the heavier
person has a larger speed at the bottom of the slide because the
larger weight causes a greater acceleration," a belief which 
did not improve much after instruction. An interviewed student
noted: {\it I remember my dad used to slide down faster because
his mass carried him down more}. Another student who chose (c)
used correct reasoning except in the last step where he did not
realize that the mass will not affect the speed: {\it \ldots
another way to see it is that your potential energy is larger than
your niece's
\ldots your kinetic energy will be larger at the bottom \ldots so
your speed is larger}. If it were a numerical problem, there is a
chance that this student would have solved it correctly, perhaps
without consciously noting that the mass drops out. For item 22,
interviews indicate that students who viewed this as a projectile
motion problem were very likely to get it wrong. A student who
chose (a) explained: {\it ball (1) gets there fastest so it will
not lose its speed as much \ldots and for the same reason ball (2)
will be faster than ball (3)}. Another student had something
similar to say: {\it the longer the path \ldots more chance
gravity has to slow the ball down}. A student who chose (c)
explained: {\it ball thrown vertically will slow down most because
all its motion is in the y direction \ldots the more is the
component in the x direction \ldots the more it remains
unchanged}. The student is probably confusing the time it takes to
reach a height with the speed at that height, and he is overlooking
the fact that all the three balls are at the same height. Another
student who chose (c) explained: {\it
\ldots smaller angle} [with horizontal] {\it will reduce downward
acceleration making the speed larger}. Incidentally, in item 22,
in the post-test, only 10\% from group L chose (e) indicating that
the speed of a ball at a given height depends on the mass, but
60\% indicated in item 15 that the speed of an object sliding down
the frictionless slide depends upon mass. This context dependent response suggests that students' knowledge is fragmented and their
recall is triggered by surface features rather than the underlying
physics concepts.

Items 1, 6, 8, and 20 probe understanding of work done by the
gravitational force. Response to item 1 shows the pervasive belief
that the work done by the gravitational force is path-dependent.
Part of the problem is the difference between the physical and
everyday definitions of work. In everyday language, more work is
considered to be done if the person becomes more tired afterward.
Item 6 can most easily be answered by invoking the definition of
work as the scalar product of force and displacement and realizing
that because the angle between the force and displacement is
$90^\circ$, the work must be zero. It can also be answered by
using the work-kinetic energy theorem. Student response to item 6
shows that instruction was more successful in helping group H
students understand the difference between velocity and speed than
group L students. Item 8 assesses students' ability to
differentiate between an applied force and the work done by it.
Many incorrectly reasoned that because a smaller magnitude tension
force is required along an inclined plane, the magnitude of the
work done by tension or the gravitational force must therefore be
smaller. The ability to distinguish between the force and the work
done by the force did not improve at all post instruction for group
L students. As mentioned earlier, part of the problem may be
students' inability to grasp the distinction between the term work
as used in physical and in everyday language. An interviewed
student who chose (d) noted: {\it \ldots you do a lot less work in
the incline case than in the vertical case \ldots it is called
mechanical advantage or something \ldots} When asked how he would
calculate the work he added: {\it \ldots I don't know if I
remember the math \ldots the concept is clear to me though \ldots
you definitely do less work when it is inclined}. Another student
had a similar explanation: {\it in inclined case it would be
easier to pull which implies less work for you}. A third student
explained: {\it tension in case (ii) is at an angle \ldots say
$30^\circ$ so the work done will have
$cos(30^\circ)$ and it will be smaller than straight up}.
Incidentally, the number of students from group L who believed
that the work done by the gravitational force is path dependent in
item 1 is significantly larger than in item 8 and this difference in number suggests that students' knowledge is fragmented and
context-dependent.\cite{mdt}

Items 9, 12, 24, and 25 probe the understanding of work done by
non-conservative forces. In item 9, most of the improvement for
group H from the pre-test to post-test was due to the transition
from choice (b) to (a), because the students learned that a
constant speed going up the hill does not imply that the
mechanical energy is conserved. The trend is opposite for group L.
A student who chose (e) explained: {\it if you ignore the
retarding effects of friction \ldots mechanical energy will be
conserved no matter what}. Other interviewed students who chose
(e) also suggested that the retarding effect of friction was the
only force that could change the mechanical energy of the system.
Although some students may have chosen (b) because they could not
distinguish between the kinetic and mechanical energies, the
following interview excerpt shows why that option was chosen by a
student despite the knowledge that kinetic and mechanical energies
are different: {\it if she goes up at constant speed then kinetic
energy does not change \ldots that means potential energy does not
change so the mechanical energy is conserved \ldots mechanical
energy is kinetic plus potential}. The student continued when
asked to explain what the potential energy is: {\it uhh \ldots
isn't it
$mgh$?}. When asked to explain why it is not changing, the student
first paused and then added: {\it \ldots $h$ is the height \ldots
I guess $h$ does change if she goes up the hill \ldots hmm \ldots
may be that means that potential energy changes. I am confused
\ldots I thought that if the kinetic energy does not change, then
potential energy cannot change \ldots aren't the two supposed to
compensate each other
\ldots is it a realistic situation that she bikes up the hill at
constant speed or is it just an ideal case?} The student is
convinced that the mechanical energy is conserved when the bike
goes up at a constant speed, and he initially thinks that both the
kinetic and potential energies must remain unchanged. When he
confronts the fact that the potential energy is changing, instead
of reasoning that the mechanical energy must be changing if the
kinetic energy is constant, he thinks that it is probably not
realistic to bike up the hill at a constant speed. He wonders if it
is only possible in the idealized physics world. Although he
ignores the work done by the non-conservative force applied on the
pedal to keep the speed constant, his statements shed light on
student's epistemological beliefs about how much one can trust
physics to explain everyday phenomena. A student who chose (c)
provided an interesting explanation: {\it in case (1) the kinetic
energy is transferred to potential energy so the mechanical energy
is conserved and in case (2) \ldots obviously \ldots if the speed
is constant \ldots mechanical energy is conserved. Case (3) is out
because she is accelerating.} This example shows student's
inconsistent reasoning and use of the term acceleration. Actually,
there is acceleration not only in case (3) but also in case (1)
(slowing down), but the student ignores it in case (1). At the same
time, he put cases (2) and (3) in different categories although the
cyclist was pedaling in both cases. For item 12, the strongest
distractor was the belief that there is a net force in the
direction of motion.\cite{mdt} This belief was essentially
unchanged by instruction for group L while half of group H
benefited from instruction. Items 24 and 25 can both be answered
using the work-kinetic energy theorem. Item 24 has previously been
investigated in depth,\cite{alan} and students had great
difficulty in realizing that because identical constant forces are
applied over the same distance to both masses (which start from
rest), their kinetic energies are identical regardless of their
masses. Some students correctly stated that the velocity of block
A will be greater, but they had difficulty in reasoning beyond
this. Interviews show that the choice (b) was often dictated by
the fact that the kinetic energy increases as the square of the
speed but only linearly with mass.\cite{alan} In item 25, all
distractors are dimensionally {\it incorrect} and equally popular. 

Items 13 and 17 probe understanding of the conservation of
mechanical energy and related issues. In item 13, some alternative
choices used energy concepts while others used momentum concepts
to assess how well students recognize the relevant concepts. In an
interview, a student explained: {\it (c) and (e) are obviously not
right because they are the same \ldots momentum and kinetic energy
both depend on mass and speed \ldots I think it is (b) because
this one falls from a greater height}. The student has forgotten
about the vector character of momentum. Interviews show that at
least some students who chose (a) were confident that the kinetic
energy has a direction (is a vector) and the kinetic energies of
the two carts cancel each other. In item 17, students who chose (d)
believed that the work done by the gravitational force on the ball
falling from the tower is negative. Interviews show that many
students did not invoke physics principles to come to this
conclusion (for example, the basic definition of work), but
thought that the work must be negative if the ball is falling in
the ``negative y direction.'' Students who
focused on speed rather than kinetic energy were likely to get
confused between options (a) and (b). An interviewed student who
chose (a) started with the following correct statement: {\it Isn't
it true that the velocity of the ball increases by like 9.8\,m/s
every second
\ldots kinetic energy is $(1/2) m v^2$} (writes down the formula)
. But later
he got misled due to a faulty proportional reasoning and added
referring to the formula: {\it $v$ increases by equal amount over
equal times\ldots so $v^2$ increases by equal amount over equal
times
\ldots mass $m$ is not changing \ldots} 

\medskip{\it Questions related to momentum concepts}.
Item 5 probes the understanding of momentum conservation in an
elastic and inelastic collision. In item 5, students choosing the
popular distractor (b) believed that the block in which the bullet
gets embedded moves faster because the bullet transfers all of its
kinetic energy to the block in the inelastic collision. The
increased frequency of distractor (a) from pre-test to post-test
shows that instruction was not effective in helping students learn
that momentum is a vector quantity and when the bullet's direction
changes during the elastic collision, it transfers momentum to the
block. Another popular distractor (c), which has several
buzz-words such as ``Newton's second law" and ``effective mass,''
is meaningless but was designed based upon student interviews.

Items 3 and 11 were designed partly to evaluate whether students
can identify the appropriate system for which momentum (and
also the kinetic energy for elastic collisions) is conserved
and indicated that many students were confused about it. Many believed that the momentum and kinetic energy
are conserved for each object. Incidentally, in the pre-test
response to item 11, 48\% noted that a person standing on ice will
remain stationary when hit by a horizontally moving ball (regardless of whether it
bounced elastically or not), because those choices had several
buzz-words.

Items 10 and 21 probe understanding of momentum conservation in an
inelastic collision or explosion. In item 10, many students
believed that the kinetic energy is conserved in an explosion.
Interviews show that some students believed that the kinetic
energy is a vector. A student who chose (a) noted: {\it it does not
lose kinetic energy because there is no friction}. Item 14 is
about the comparison of two completely inelastic collisions, but
the alternative choices included both the energy and momentum
concepts and students had to determine which was appropriate. Even
in item 21, different alternative choices included both momentum
and mechanical energy conservation to assess if students could
identify the relevant concepts. An interviewed student noted: {\it
I know this is an inelastic collision \ldots I will pick (a)
because rain is falling vertically and is not going to influence
the speed of the cart \ldots}. When asked about his thoughts on
choice (c), the student paused and then added: {\it 
\ldots probably the horizontal momentum is conserved but the rain
did not have any horizontal momentum so the velocity of the cart in
the horizontal direction is unchanged \ldots that's why I picked
(a) \ldots} The student's last statement is inconsistent because if
the horizontal momentum is conserved, then the addition of mass
should slow the cart. It is unclear however what the student meant
by ``the horizontal momentum is conserved." Taking item 3 as a
guide, the student may very well be just referring to the cart as
the system for which the horizontal momentum is conserved, although
choice (c) explicitly noted the ``cart-rain" system.

Items 19 and 23 are applications of the impulse-momentum theorem.
In item 19, students had to realize that the change in momentum is
the same whether the biker slams into a haystack or a concrete
wall, but the haystack changes the momentum over a longer time and
reduces the average force. In the interview, a student who chose
(a) did not use the appropriate physics principle as the starting
point and explained: {\it \ldots haystack gives you a smaller
force so the impulse is smaller}. A student who chose (e) noted:
{\it \ldots harder objects give up more energy than softer
objects}. In item 23, the change in momentum of the rubber ball in
the elastic collision with the surface is twice compared to that
of the putty ball in the completely inelastic collision. Because
the time
$\Delta t$ over which the velocity of the balls changes from its
initial to its final value due to the contact with the surface is
the same, the average force on the surface by the rubber ball is
twice as much as the putty ball. Interviews show that many
students who chose distractor (a) reasoned that because the masses
are the same, the force is proportional to the speed. Because both
balls have the same speed when they hit the surface, the average
force exerted on the surface is the same, regardless of whether the
ball bounces back or gets stuck. Some students focused on the
weights of the two balls and said that the balls would exert the
same force on the surface because their weights were the same. Some
said they did not know why $\Delta t$ was provided, because the
weight does not depend upon time.

After instruction, students' overall performance improved on all
test items (see gains in Table~I) except item 16 on the ballistic
pendulum, which was the only one that involved both the energy
and momentum concepts simultaneously. Interviews show that some
students who verbalized that the problem had several parts often
ended up focusing on only one aspect. One student who chose (b)
explained: {\it because they are sticking \ldots it is a collision
\ldots so momentum is conserved.} Poor performance on this item,
which involved both the energy and momentum concepts, suggests that
even group H had difficulty analyzing multi-part problems
involving more than one concept\cite{singh2}. In the post-test, student
performance on items 7 and 18, which focused more on the
definition and factual information (lower cognitive complexity),
was above average. Responding to item 7, an interviewed student
gave the following reasoning for why the motorcycle has a larger
momentum: {\it
\ldots it has an acceleration which contributes to force which is
involved in momentum}.

\section{Summary}
We have developed a research-based test to assess introductory
students' conceptual understanding of energy and momentum. We found
that students have difficulty in qualitatively interpreting the
basic principles related to energy and momentum and in applying
them in physical situations. The difficulties were not strongly
dependent on student population or calculus background, except for
honors and active-engagement classes. Comparison of the
pre-/post-test scores of students in the lower (L) and upper (H)
quartiles based upon the overall test score shows that the
upper-quartile group had significantly higher normalized gains.
The test can be administered as a pre-/post-test to assess the
difficulties prior to instruction, as well as those that remain
uncorrected following a particular type of instructional
intervention. The test can be used to compare the understanding of
energy and momentum concepts in courses employing different
instructional designs and strategies. 

\begin{acknowledgments}
We are very grateful to all the faculty who reviewed the various
components of the test at several stages and provided invaluable
feedback. We are also very thankful to all the faculty who
administered the test. This work is supported in part by the
National Science Foundation and the Spencer Foundation.
\end{acknowledgments}

\appendix
\section{Energy and Momentum Conceptual Survey}\label{appa}

\begin{center}
{\bf Instructions} 
\end{center}
$\bullet$ Select \underline{one} of the five choices (a)--(e) for each of the 25
questions.\\
$\bullet$ Ignore the retarding effects of friction and air resistance unless
otherwise stated.

\begin{enumerate}

\item You lift a suitcase from the floor to a table. In addition to
the weight of the suitcase, select all of the following factors
that determine the work done by the gravitational force on the
suitcase.

\begin{enumerate}[(1)]

\item whether you lift it directly up to the table or
along a longer path

\item whether you lift it quickly or slowly

\item the height of the table above the floor

\end{enumerate}

\begin{enumerate}[{\bf (a)}]

\item (1) only
\item (3) only
\item (1) and (3) only
\item (2) and (3) only
\item (1), (2) and (3)

\end{enumerate}

\item Two frictionless slides are shaped differently but start at
the same height $h$ and end at the same level as shown below. You
and your friend, who has the same weight as you, slide down from the
top on different slides starting from rest. Which one of the
following statements best describes who has a {larger} speed at the
bottom of the slide?
\begin{center}
\epsfig{file=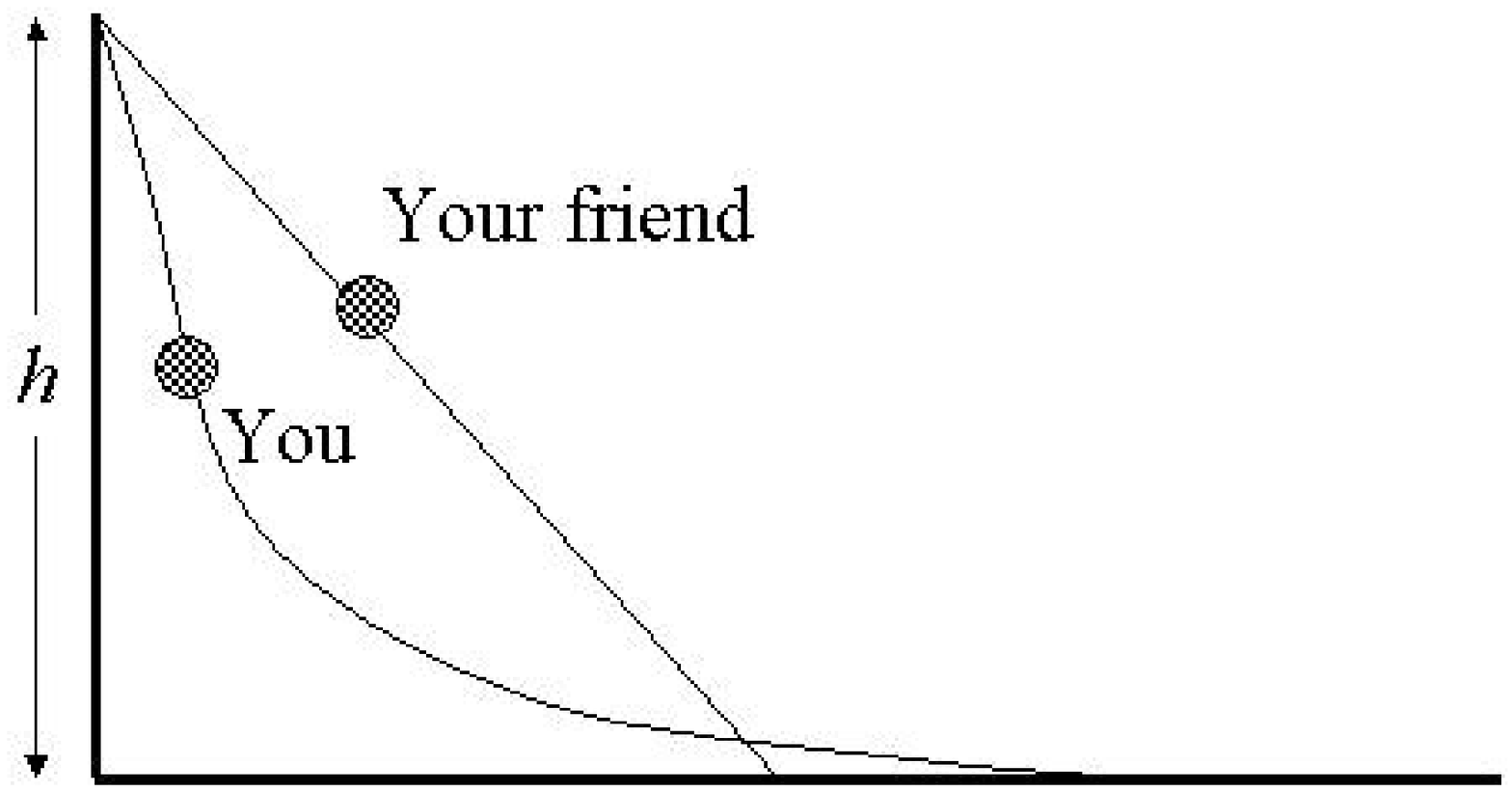,height=1.9in}
\end{center}
\begin{enumerate}[(1)]
\item You, because you initially encounter a steeper slope so that
there is more opportunity for accelerating.

\item You, because you travel a longer distance so that there is more
opportunity for accelerating.

\item Your friend, because her slide has a constant slope so that
she has more opportunity for accelerating.

\item Your friend, because she travels a shorter distance so that
she can conserve her kinetic energy better.

\item Both of you have the same speed.

\end{enumerate}

\item A moving white hockey puck collides \underline{elastically}
with a stationary red hockey puck on a frictionless horizontal
surface. No net external force acts on the two-puck system. Select
all of the following statements that {must} be \underline{true} for
this elastic collision.

\begin{enumerate}[(1)]

\item The kinetic energy of the
{white puck} is conserved (same before and after the collision).

\item The linear momentum of the {white puck} is conserved.

\item The linear momentum of the {two-puck} system is conserved.

\end{enumerate}

\begin{enumerate}[{\bf (a)}]

\item (1) only
\item (3) only
\item (1) and (2) only
\item (1) and (3) only
\item (1), (2) and (3)
\end{enumerate}

\item Two identical stones, A and B, are shot from a cliff from the
\underline{same height} and with 
\underline{identical initial speeds $v_0$}. Stone A is shot
vertically up, and stone B is shot vertically down (see Figure). 
Which one of the following statements best describes which stone has
a larger speed {right before} it hits the ground

\begin{center}
\epsfig{file=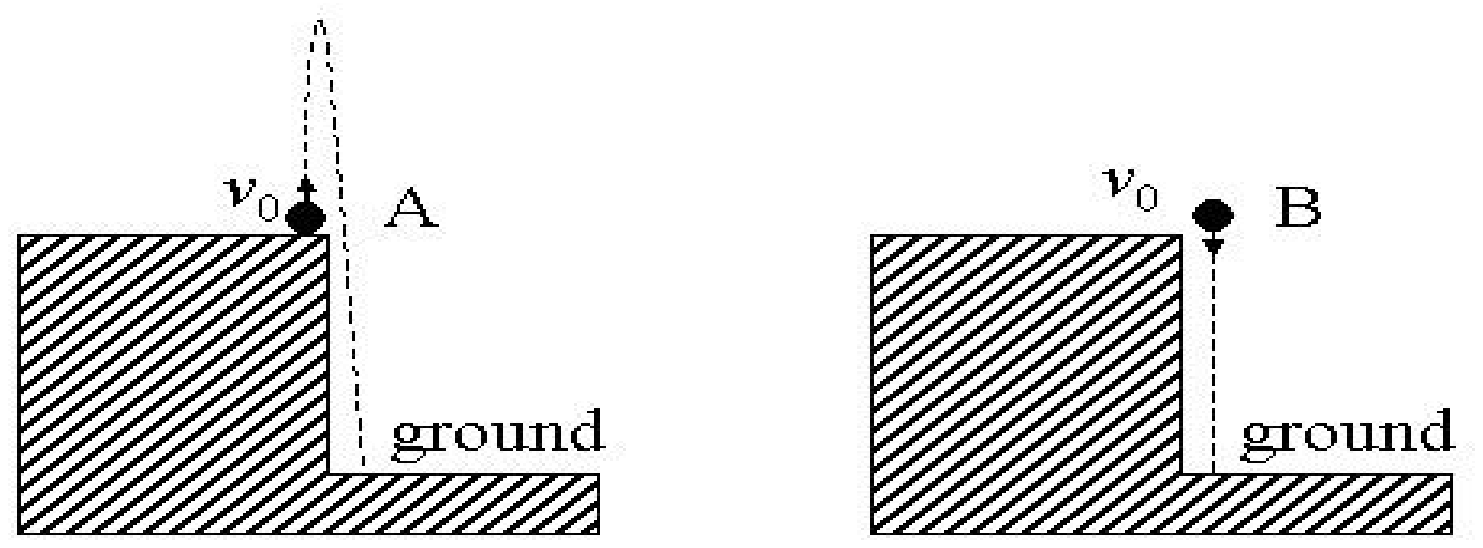,height=1.9in}
\end{center}

\begin{enumerate}[{\bf (a)}]

\item Both stones have the same speed.

\item A, because it travels a longer path.
\item A, because it takes a longer time.

\item A, because it travels a longer path and takes a longer time.

\item B, because no work is done against the gravitational force.
\end{enumerate}

\item
Two \underline{identical} bullets are fired horizontally with
\underline{identical speeds $v_0$} at two blocks of
\underline{equal mass}. The blocks rest on a frictionless horizontal
surface and are made of hard steel and soft wood respectively (see
Figure). One bullet bounces
{elastically} off the steel block. The other bullet becomes embedded
inside the wood block. Which one of the following statements best
describes which block travels faster after the collision?

\begin{center}
\epsfig{file=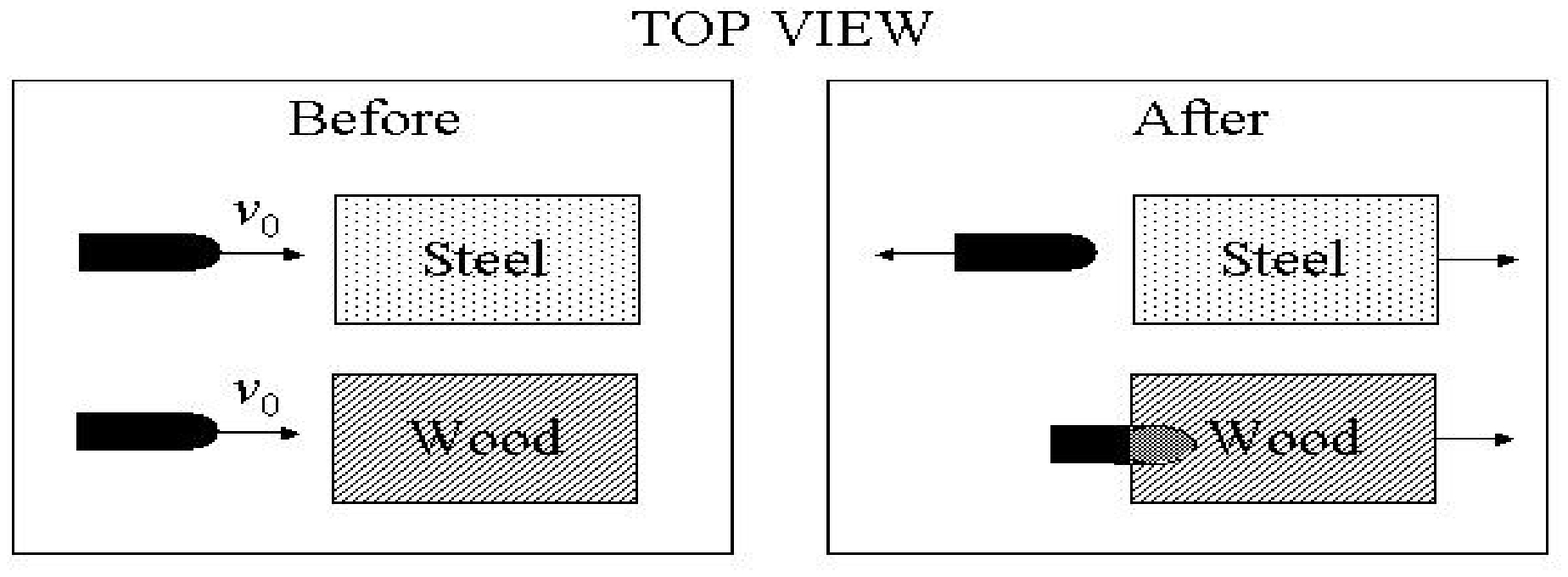,height=1.7in}
\end{center}

\begin{enumerate}[{\bf (a)}]

\item The wood block, because it has gained the momentum of the
bullet, while the other bullet does not impart its momentum to the
steel block.

\item The wood block, because the bullet transfers all of its
kinetic energy to it.

\item The wood block, because its larger effective mass after the
collision, in accordance with Newton's second law, results in a
larger force to accelerate the block.

\item The steel block, because the bullet bounces off from it.

\item Both blocks travel with the
same speed.
\end{enumerate}

\item A satellite is moving around Earth in a \underline{circular}
orbit at a \underline{constant speed} (see Figure). The only force
that acts on the satellite is Earth's gravitational force which
points directly toward earth's center. Which one of the following
statements is \underline{true} as the satellite moves
\underline{from point A to point B} in the orbit?

\begin{center}
\epsfig{file=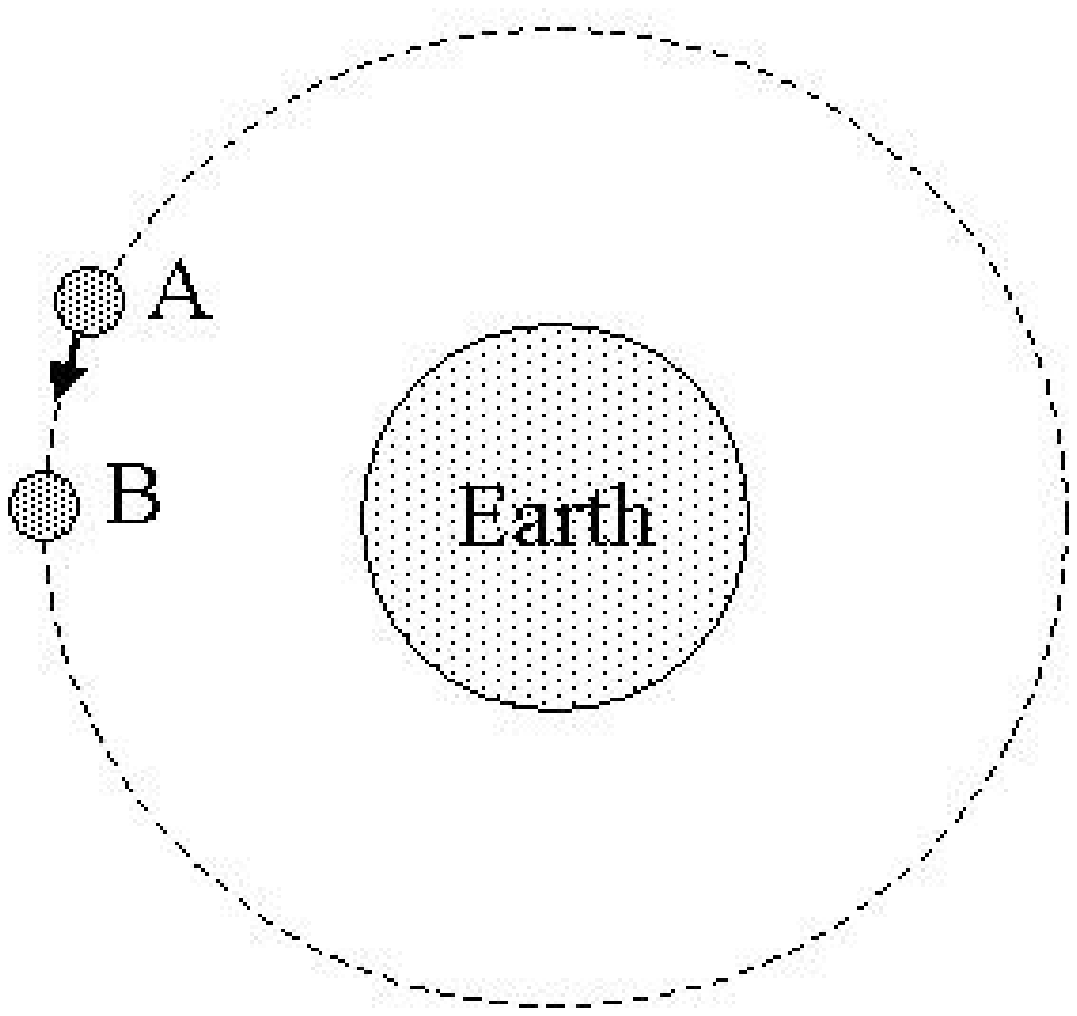,height=1.8in}
\end{center}

\begin{enumerate}[{\bf (a)}]

\item The gravitational potential energy of the satellite decreases
as it moves from A to B.

\item The work done on the satellite by the gravitational force is
negative for the motion from A to B.

\item The work done on the satellite by the gravitational force is
zero for the motion from A to B.

\item The velocity of the satellite remains unchanged as it moves from
A to B.

\item None of the above.

\end{enumerate}

\item A motorcycle and a truck are both moving in the {same direction}
in adjacent lanes on a highway. At a particular instant, the speed of
the motorcycle is four times as large as the speed of the truck $v$. 
At that instant, the motorcycle is accelerating forward while the
truck is moving at a constant velocity. Which one of the following
statements best describes which vehicle has a \underline{larger
momentum} at that instant?

\begin{center}
\epsfig{file=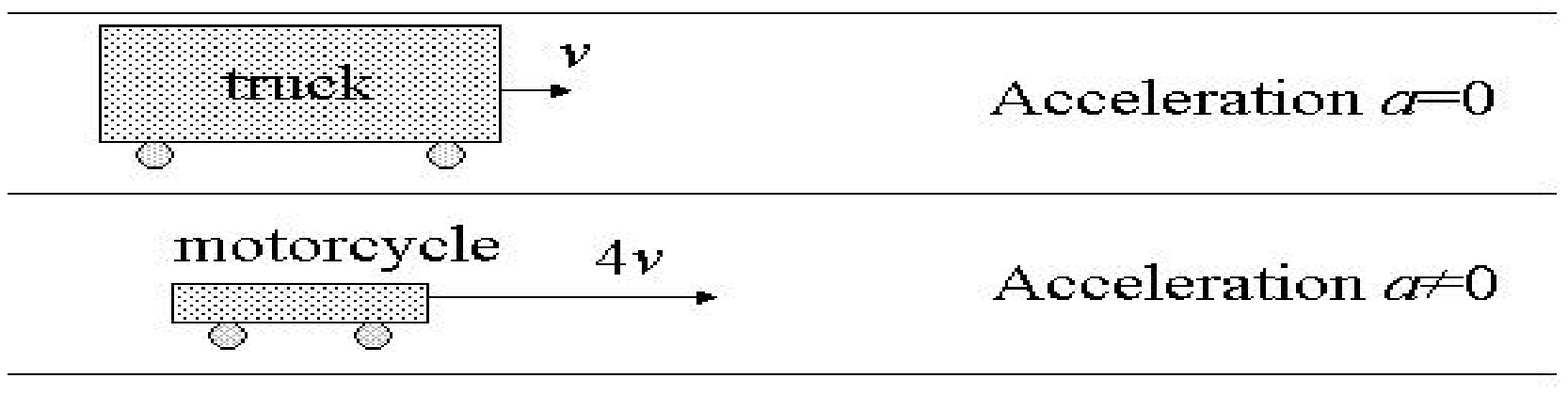,height=1.4in}
\end{center}

\begin{enumerate}[{\bf (a)}]

\item The truck, because it has a larger mass.

\item The motorcycle, because it is moving faster.

\item The motorcycle, because it has an acceleration.

\item The motorcycle, because it is moving faster and also has an
acceleration.

\item Not enough information.
\end{enumerate}

\item You want to lift a heavy block through a height $h$ by
attaching a string of negligible mass to it and pulling so that it
moves at a 
\underline{constant velocity}. You have the choice of lifting it
either by pulling the string vertically upward or along a
{frictionless} inclined plane (see Figure). Which one of the
following statements is \underline{true}?
\begin{center}
\epsfig{file=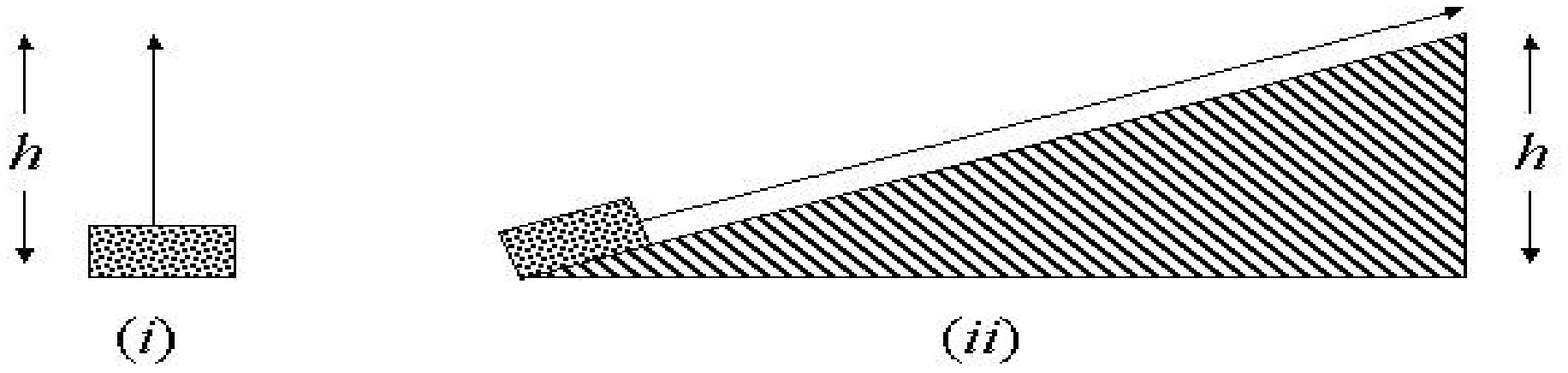,height=1.8in}
\end{center}

\begin{enumerate}[{\bf (a)}]

\item The magnitude of the tension force in the string is smaller in
case (i) than in case (ii).

\item The magnitude of the tension
force in the string is the same in both cases.

\item The work
done on the block by the tension force is the same in both cases.

\item The work done on the block by the tension force is smaller
in case (ii) than in case (i).

\item The work done on the block
by the gravitational force is smaller in case (ii) than in
case (i).
\end{enumerate}

\item
Three bicycles approach a hill as described below:

\begin{enumerate}[(1)]
\item Cyclist 1 stops pedaling at the bottom of the hill, and her bicycle 
 coasts up the hill.

\item Cyclist 2 pedals so that her bicycle goes up the hill at a constant speed.

\item Cyclist 3 pedals harder, so that her bicycle accelerates up the hill.
\end{enumerate}

Ignoring the retarding effects of friction, select all the cases in which the
\underline{total mechanical energy} of the cyclist and bicycle is conserved.

\begin{enumerate}[{\bf (a)}]

\item (1) only
\item (2) only
\item (1) and (2) only
\item (2) and (3) only
\item (1), (2) and (3)
\end{enumerate}

\item A bomb at rest on a \underline{horizontal frictionless} surface
explodes and breaks into three pieces that fly apart
\underline{horizontally} as shown below. Select all of the following
statements that must be \underline{true} after the bomb has
exploded.

\begin{enumerate}[(1)]
\item The total kinetic energy of the bomb fragments is the same as that of the
bomb before explosion.

\item The total momentum of the bomb fragments is the same as that
 of the bomb before explosion.

\item The total momentum of the bomb fragments is zero.

\end{enumerate}

\begin{center}
\epsfig{file=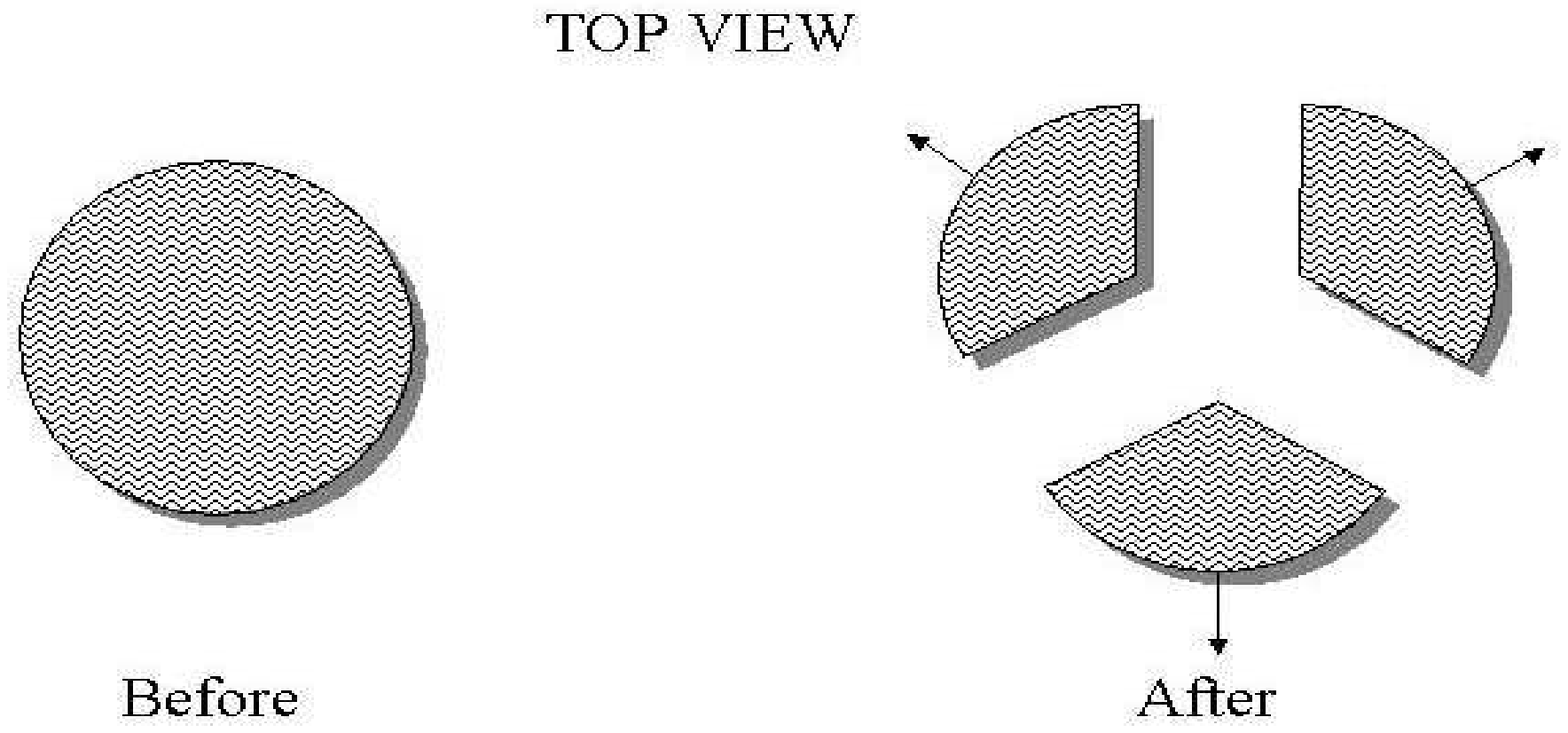,height=2.1in}
\end{center}

\begin{enumerate}[{\bf (a)}]
\item (1) only
\item (2) only
\item (3) only
\item (2) and (3) only
\item (1), (2) and (3)
\end{enumerate}

\item You and your friend are both standing on a horizontal
frictionless surface. To get your friend's attention, you throw a
ball due west at your friend as shown below. The ball bounces
elastically off your friend's back. Which one of the following
statements is \underline{true} about this situation?

\begin{center}
\epsfig{file=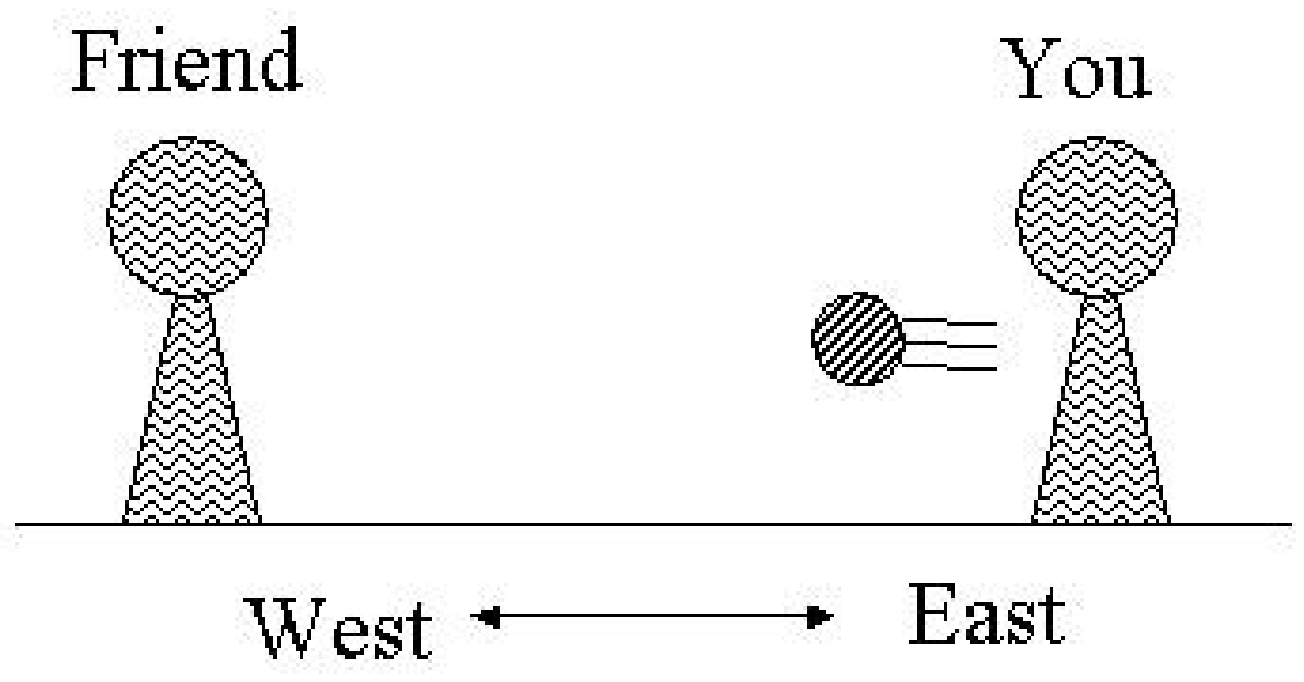,height=1.5in}
\end{center}

\begin{enumerate}[{\bf (a)}]
\item Your friend will remain stationary because the ball bounces
elastically and does not impart its momentum to her.

\item Your friend will remain stationary because the kinetic
energy of the ball is conserved 
in an elastic collision.

\item Your friend will remain stationary due to conservation of
both linear momentum 
and kinetic energy.

\item You will remain stationary after you throw the ball due to
conservation of both
linear momentum and kinetic energy.

\item You will move east after you throw the ball due to
conservation of linear momentum.
\end{enumerate}

\item Using a rope of negligible mass, you pull a box along a
horizontal surface with a constant horizontal force ${\bf F_A}$. 
The box moves at a \underline{constant velocity} from position A to
position B. The force of friction ${\bf F_k}$ 
\underline{cannot} be neglected. Which one of the following
statements concerning the motion of the box from A to B is
\underline{true}?

\begin{center}
\epsfig{file=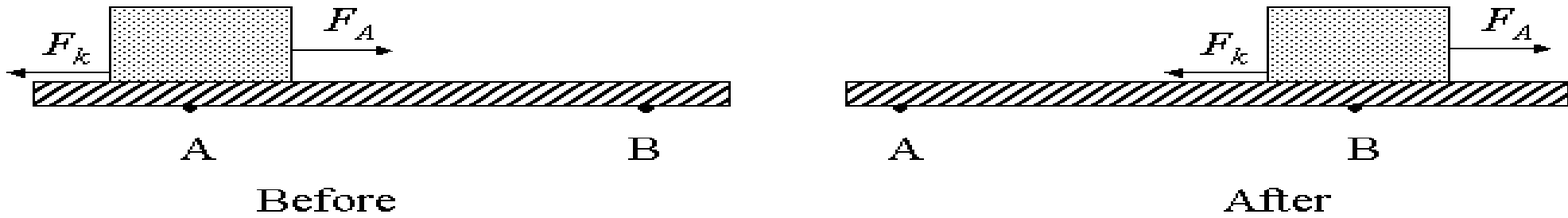,height=1.1in}
\end{center}

\begin{enumerate}[{\bf (a)}]

\item The work done on the box by the gravitational force is
non-zero.

\item The work done on the box by ${\bf F_k}$ is positive.

\item The total work done on the box by the net force is non-zero.

\item The magnitude of the work done on the box by ${\bf F_A}$ is
equal to the magnitude of the work done by ${\bf F_k}$.

\item The magnitude of ${\bf F_A}$ is greater than the magnitude of
${\bf F_k}$.

\end{enumerate}

{\bf {\Large $\bullet$} \underline{Setup for Questions (13) and
(14)}}\\

Carts A and B are identical in all respects before the collision.\\

{\bf Figure (i)}: Cart A starts from rest on a hill at a height $h$ above the ground. It rolls 
down and collides ``head-on" with cart B which is initially at rest on the ground. 
The two carts stick together. 

{\bf Figure (ii)}: Carts A and B are at rest on opposite hills at heights $h/2$
above the ground. They roll down, collide ``head-on" with each other on the ground
and stick together.

\begin{center}
\epsfig{file=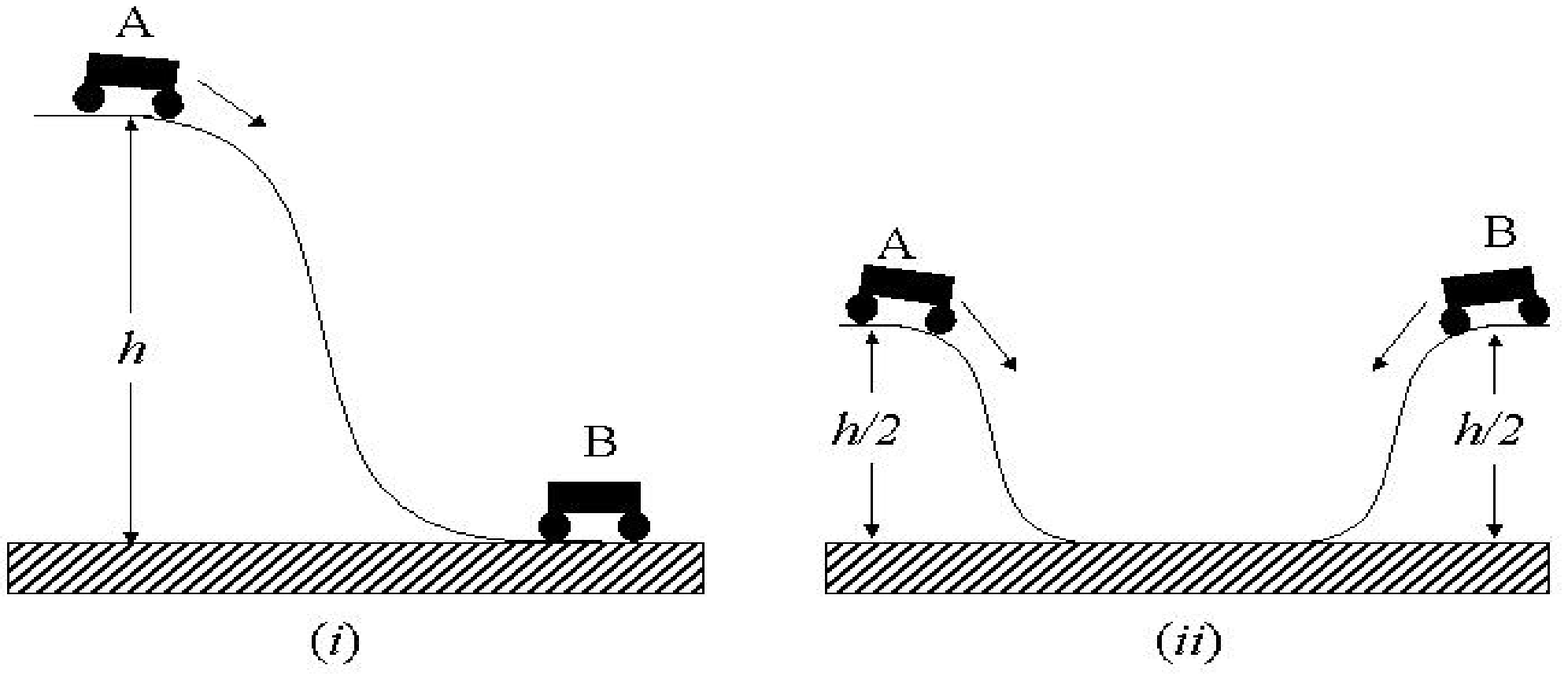,height=1.9in}
\end{center}

\item
Which one of the following statements is \underline{true} about the
\underline{two-cart system} {just before} the carts collide in the two cases? \\

\underline{Just before} the collision on the ground,
\begin{enumerate}[{\bf (a)}]

\item the kinetic energy of the system is zero in case (ii).

\item the kinetic energy of the system is greater in case (i) than in
case (ii).

\item the kinetic energy of the system is the same
in both cases.

\item the momentum of the system is greater in
case (ii) than in case (i).

\item the momentum of the system is
the same in both cases.
\end{enumerate}

\item
Which one of the following statements is \underline{true} about the \underline{two-cart system} {just after} the carts collide 
in the two cases? \\

\underline{Just after} the collision,
\begin{enumerate}[{\bf (a)}]
\item the kinetic energy of the system is greater in case (ii) than
in case (i).

\item the kinetic energy of the system is the same
in both cases.

\item the momentum of the system is greater in
case (ii) than in case (i).

\item the momentum of the system is
non-zero in case (i) while it is zero in case (ii).

\item the
momentum of the system is the same in both cases.
\end{enumerate}

\item While in a playground, you and your niece take turns sliding
down a frictionless slide. Your mass is 75 kg while your little
niece's mass is only 25 kg. Assume that both of you begin sliding
\underline{from rest from the same height}. Which one of the
following statements best describes who has a {larger} speed at the
bottom of the slide?

\begin{enumerate}[{\bf (a)}]

\item Both of you have the same
speed at the bottom.

\item Your niece, because she is not
pressing down against the slide as strongly so her motion
is closer to free fall than yours.

\item You, because your greater weight causes a greater downward
acceleration.

\item Your niece, because lighter objects are
easier to accelerate.

\item You, because you take less time to
slide down.
\end{enumerate}

\item
Two small spheres of putty, A and B, of equal mass, hang from the
ceiling on massless strings of equal length. Sphere A is raised to a
height $h_0$ as shown below and released. It collides with sphere B
(which is initially at rest); they stick and swing together to a
maximum height $h_f$. The height $h_f$ can be determined in terms of
$h_0$ by invoking which of the following principles:

\begin{enumerate}[(1)]
\item the conservation of mechanical energy
\item the conservation of linear momentum 
\end{enumerate}

\begin{center}
\epsfig{file=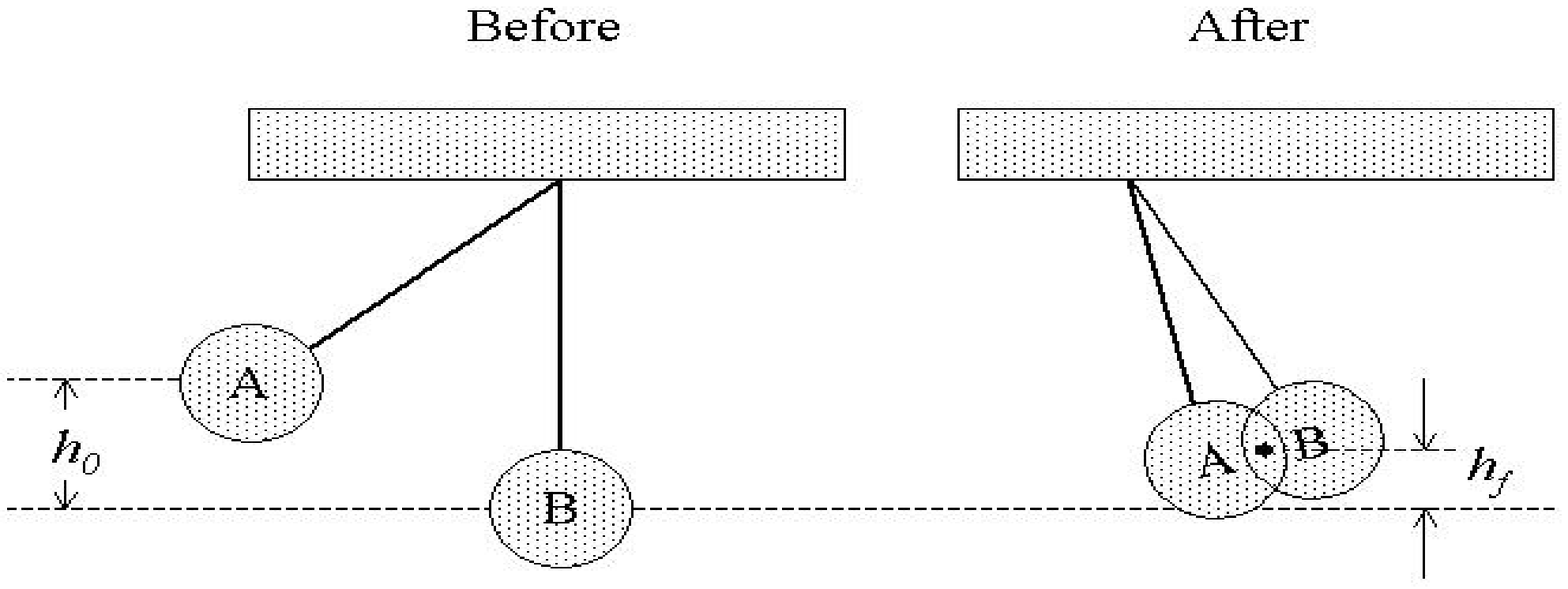,height=1.9in}
\end{center}

\begin{enumerate}[{\bf (a)}]
\item (1) only
\item (2) only
\item both (1) and (2)
\item either (1) or (2) but not both
\item Principles (1) and (2) alone are not sufficient to find $h_f$
in terms of $h_0$.
\end{enumerate}

\item
You drop a ball from a high tower and it falls freely under the influence of the gravitational force.
Which one of the following statements is \underline{true}?

\begin{enumerate}[{\bf (a)}]
\item The kinetic energy of the ball increases by equal amounts in
equal times.

\item The kinetic energy of the ball increases by equal amounts over
equal distances.

\item There is zero work done on the ball by the gravitational force
as it falls.

\item The work done on the ball by the gravitational force is
negative as it falls.

\item The total mechanical energy of the ball decreases as it falls.

\end{enumerate}

{\bf {\Large $\bullet$} {In Questions (18) and (19) below, the \underline{impulse} of
a force is defined as the \underline{product} of the average force and the time
interval during which the force acts.}}

\item Which one of the following statements is \underline{true}
concerning linear momentum?

\begin{enumerate}[{\bf (a)}]
\item Momentum is a force.
\item The momentum of an object is always positive.
\item Momentum is a scalar quantity.
\item The SI unit of momentum is kg\,m$^2$/s.
\item Momentum and impulse have the same units.
\end{enumerate}

\item The brakes of your bicycle have failed, and you must choose
between slamming into either a haystack or a concrete wall. Which one
of the following statements \underline{best justifies} why hitting a
haystack is a wiser choice than hitting a concrete wall?

\begin{enumerate}[{\bf (a)}]
\item The haystack gives you a smaller impulse than the concrete
wall.

\item The haystack changes your momentum over a longer time.

\item Your change in kinetic energy is smaller if you hit the
haystack than if you hit the concrete wall.

\item Your change in momentum is smaller if you hit the haystack
than if you hit the concrete wall.

\item More potential energy is stored in the wall which is
released upon the impact and
results in a greater force on you.
\end{enumerate}

\item You slide down two consecutive slopes of frictionless ice whose
vertical heights $h$ are identical, as shown below. Select all of the
following statements that must be \underline{true}.

\begin{enumerate}[(1)]
\item The change in your kinetic energy is \underline{identical} for the motion
from A to B and from B to C.

\item The work done on you by the gravitational force is \underline{smaller} for the
motion from A to B than from B to C.

\item The work done on you by the gravitational force is \underline{greater} for
the motion from A to B than from B to C.
\end{enumerate}

\begin{floatingfigure}[r]{3.4in}
\epsfig{file=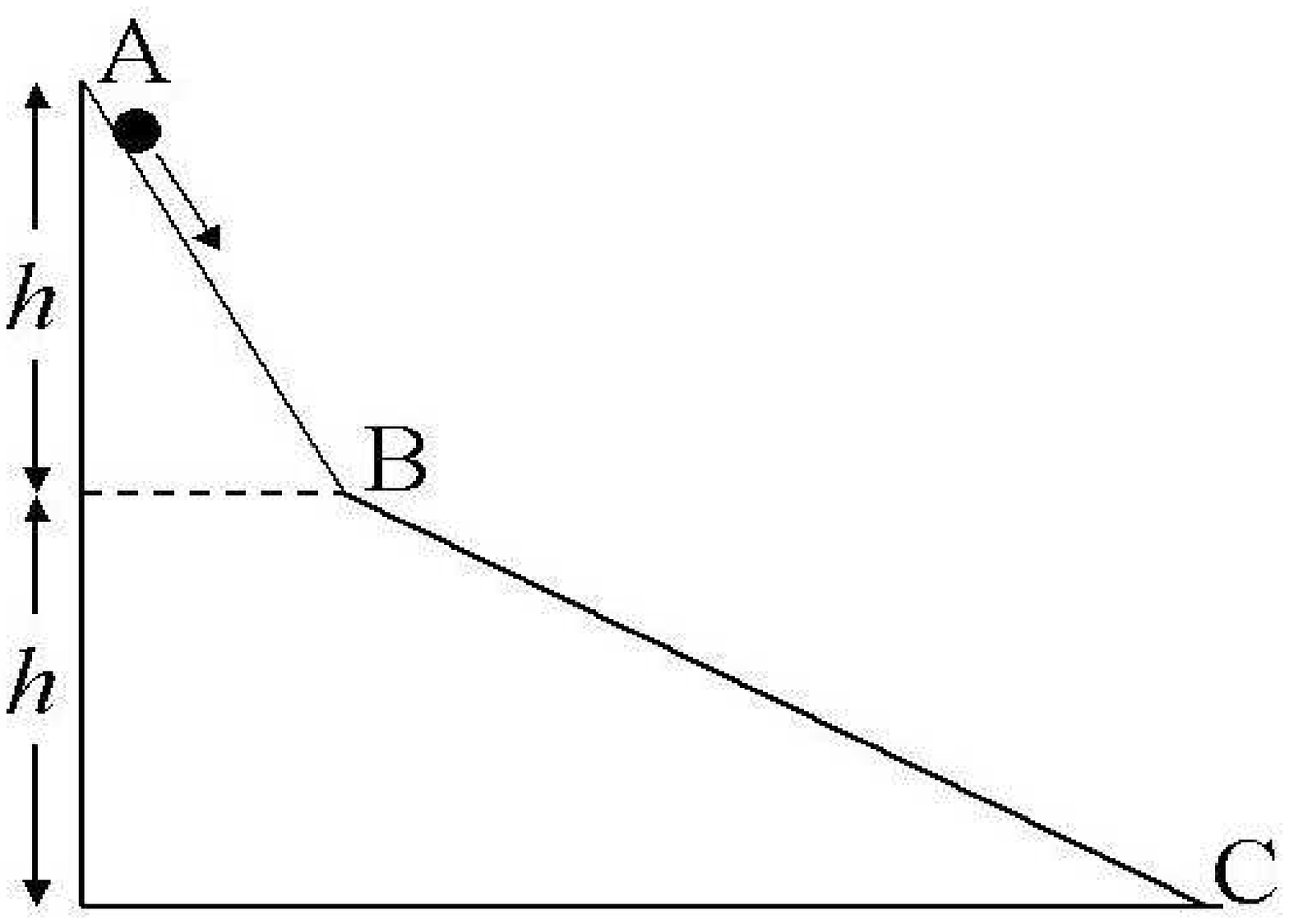,height=1.8in}
\end{floatingfigure}

\begin{enumerate}[{\bf (a)}]
\item (1) only
\item (2) only
\item (3) only
\item (1) and (2) only
\item (1) and (3) only
\end{enumerate}

\item Rain starts falling vertically down into a cart with
frictionless wheels which is \underline{initially} moving at a
constant velocity on a horizontal surface. The rain drops come to
rest with respect to the cart after striking it, and rain water
accumulates in the cart. Select all of the following statements that
must be \underline{true} about this situation.

\begin{enumerate}[(1)]
\item The cart will continue to move at a constant velocity
because the rain is falling vertically while the cart is moving
horizontally.

\item The cart will continue to move at a constant velocity
because the total mechanical energy of the cart-rain system is
conserved.

\item The cart will slow down because the horizontal momentum of
the cart-rain system is conserved.
\end{enumerate}

\begin{floatingfigure}[r]{3.4in}
\epsfig{file=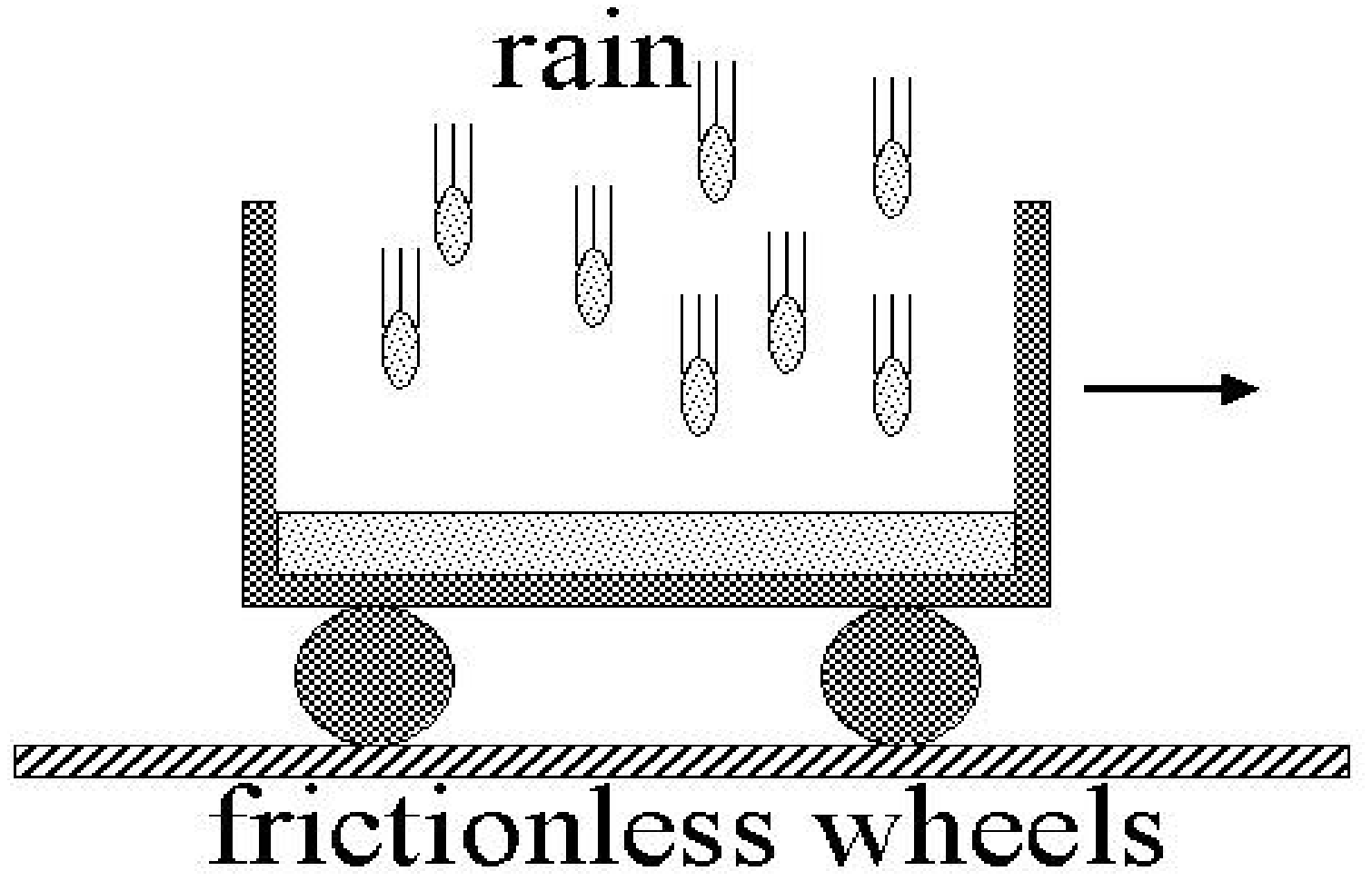,height=1.8in}
\end{floatingfigure}

\begin{enumerate}[{\bf (a)}]
\item (1) only
\item (2) only
\item (3) only
\item (1) and (2) only
\item None of the above
\end{enumerate}

\item Three balls are launched from the \underline{same} horizontal
level with \underline{identical speeds $v_0$} as shown below. Ball
(1) is launched vertically upward, ball (2) at an angle of
$60^\circ$, and ball (3) at an angle of $45^\circ$. In order of
{decreasing speed} (fastest first), rank the \underline{speed}
each one attains when it reaches the level of the
\underline{dashed horizontal line}. All three balls have
sufficient speed to reach the dashed line.

\begin{center}
\epsfig{file=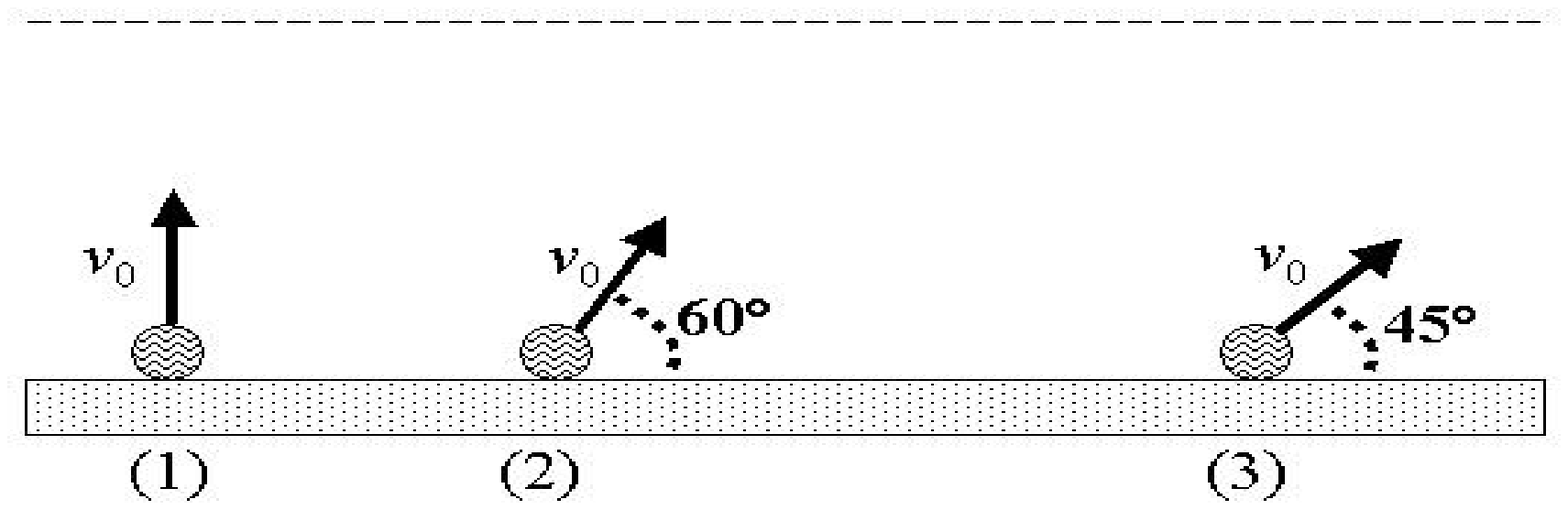,height=1.4in}
\end{center}

\begin{enumerate}[{\bf (a)}]
\item (1), (2), (3)
\item (1), (3), (2)
\item (3), (2), (1)
\item They all have the same speed.
\item Not enough information, their speeds will depend on their
masses.
\end{enumerate}

\item You drop two balls of \underline{equal mass}, made of rubber
and putty, from the \underline{same height $h$} above a horizontal
surface (see Figure). The rubber ball bounces up after it strikes
the surface while the putty ball comes to rest after striking it.
Assume that in both cases the velocity of the ball takes the
\underline{same time $\Delta t$} to change from its initial to its
final value due to contact with the surface. Compare the
\underline{average forces} $\overline {\bf F}_R$ and $\overline {\bf
F}_P$ exerted on the surface by the rubber and putty balls,
respectively, \underline{during time $\Delta t$}.

\begin{center}
\epsfig{file=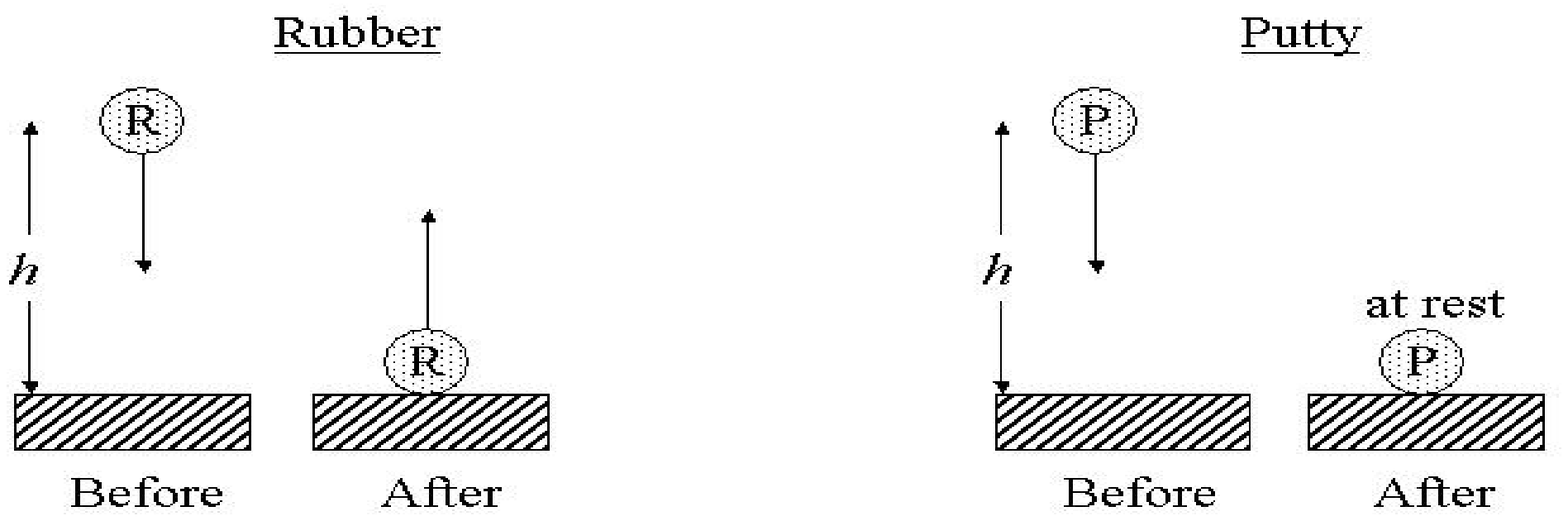,height=1.7in}
\end{center}

\begin{enumerate}[{\bf (a)}]
\item $\overline {\bf F}_R= \overline {\bf F}_P$
\item $ \overline {\bf F}_R > \overline {\bf F}_P$
\item $ \overline {\bf F}_R < \overline {\bf F}_P$
\item $\overline {\bf F}_R$ may be smaller or larger than $\overline
{\bf F}_P$ depending upon the relative size of the balls.

\item 
$\overline {\bf F}_R$ may be smaller or larger than $\overline {\bf
F}_P$ depending upon the actual height $h$ from which the balls are
dropped.
\end{enumerate}

\item Two blocks are initially at rest on a {frictionless horizontal}
surface. The mass $m_A$ of block A is \underline{less} than the mass
$m_B$ of block B. You apply the \underline{same constant force ${\bf
F}$} and pull the blocks through the \underline{same distance $d$}
along a straight line as shown below (force ${\bf F}$ is applied for
the entire distance $d$). 

\begin{center}
\epsfig{file=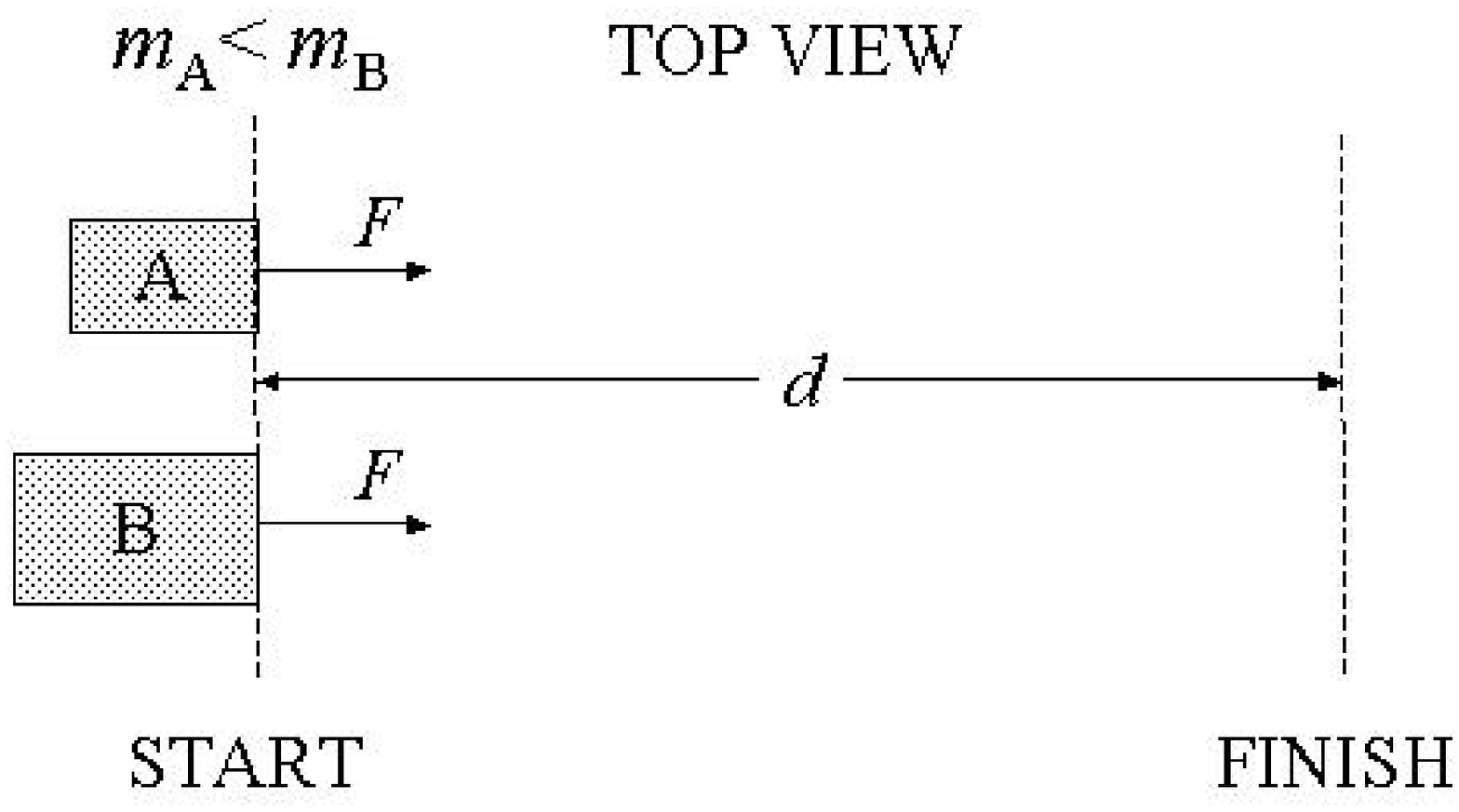,height=1.9in}
\end{center}

Which one of the following statements correctly compares the
\underline{kinetic energies} of the blocks after you pull them the 
\underline{same distance $d$}?

\begin{enumerate}[{\bf (a)}]
\item The kinetic energies of both blocks are identical.

\item The kinetic energy is greater for the smaller mass block
because it achieves a larger 
speed.

\item The kinetic energy is greater for the larger mass block
because of its larger mass.

\item Not enough information, need to know the actual mass of
both blocks to compare the kinetic energies.

\item Not enough information, need to know the actual magnitude
of force ${\bf F}$ to compare the 
kinetic energies.
\end{enumerate}

\item A box slides with an initial speed $v_o$ on a horizontal
surface \underline{with friction} and eventually comes to a stop. 
Which one of the following is equal to the {change in the kinetic
energy} of the box?

\begin{center}
\epsfig{file=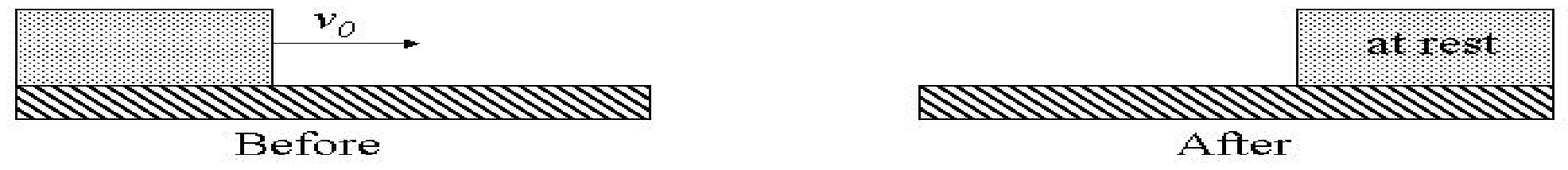,height=1.2in}
\end{center}

\begin{enumerate}[{\bf (a)}]

\item The momentum of the box multiplied by the distance
travelled before coming to rest.

\item The momentum of the box multiplied by the time elapsed
before coming to rest.

\item The momentum of the box multiplied by the deceleration of
the box.

\item The mass of the box multiplied by the deceleration of the
box.

\item None of the above.
\end{enumerate}

\end{enumerate}

\newpage

\begin{table}[h]
\centering
\begin{tabular}[t]{|c|c|c|c|c|c|c|c|c|c|}
\hline Item $\#$&Pre-H&Post-H&Pre-L&Post-L&Pre-A&Post-A &$g_H$&$g_L$&$g_A$\\
\hline
\hline 1&57 & 82 & 14 & 36 & 33 & 63 &		0.57&	0.25&	0.45\\
\hline 2&78 & 97 & 25 & 59 & 50 & 81 &		0.86&	0.45&	0.63\\
\hline 3&66 & 82 & 24 & 36 & 45 & 61 &		0.47&	0.15&	0.30\\
\hline 4&68 & 89 & 31 & 42 & 47 & 69 &		0.65&	0.16&	0.40\\
\hline 5&18 & 51 & 4 & 7 & 11 & 27 &		0.40&	0.04&	0.18\\
\hline 6& 19 & 52 & 11 & 16& 12 & 28 &	0.40&	0.06&	0.18\\
\hline 7& 75 & 97 & 24 & 62 & 55 & 85 &		0.88&	0.50&	0.66\\
\hline 8&54 & 70 & 14 & 13 & 30 & 45 &		0.35&	-0.02&	0.22\\
\hline 9&36 & 61 & 27 & 18 & 26 & 36 &		0.38&	-0.12&	0.13\\
\hline 10& 22 & 57 & 5 & 15 & 13 & 35 &		0.44&	0.10&	0.25\\
\hline 11& 68 & 95 & 14 & 56 & 41 & 79 &		0.84&	0.48&	0.65\\
\hline 12& 34 & 57 & 13 & 17 & 21 & 33 &		0.34&	0.04&	0.15\\
\hline 13& 69 & 75 & 28 & 35 & 52 & 53 &		0.18&	0.10&	0.03\\
\hline 14& 78 & 89 & 37 & 52 & 60 & 70 &		0.49&	0.24&	0.25\\
\hline 15& 74 & 80 & 24 & 40 & 47 & 58 &		0.23&	0.21&	0.22\\
\hline 16& 41 & 36 & 29 & 17 & 34 & 26	 &	-0.08&	-0.18&	-0.11\\
\hline 17& 34 & 60 & 23 & 32 & 30 & 46 &		0.39&	0.11&	0.23\\
\hline 18& 32 & 84 & 12 & 50 & 20 & 66 &		0.76&	0.43&	0.57\\
\hline 19& 71 & 82 & 22 & 52 & 47 & 70 &	 0.37&	0.39&	0.43\\
\hline 20& 62 & 80 & 11 & 30 & 35 & 58 &		0.46&	0.21&	0.34\\
\hline 21& 51 & 83 & 36 & 42 & 44 & 62	 &	0.65&	0.10&	0.32\\
\hline 22& 28 & 48 & 6 & 7 & 16 & 29 & 		0.27&	0.01&	0.16\\
\hline 23& 27 & 41 & 16 & 18 & 21 & 25 &		0.20&	0.02&	0.06\\
\hline 24& 34 & 47 & 6 & 9 & 20 & 29 &		0.19&	0.03&	0.11\\
\hline 25&47 & 75 & 20 & 44 & 33 & 59	 &	0.52&	0.30&	0.39\\
\hline
\hline
\end{tabular}
\caption{Student performance before and after instruction. The average percent pre-/post-test
scores and normalized gain $g$ on each item for the upper 25\% (H), Lower 25\%
(L), and for all students (A).}
\label{table1}
\end{table}

\begin{table}[h]
\centering
\begin{tabular}[t]{|c|c|c|c|c|c||c|c|c|c|c||c|c|c|c|c||c|c|c|c|c|}
\hline Item &\multicolumn{5}{|c|}{pre-test (L)}&\multicolumn{5}{|c|}{post-test
(L)}&\multicolumn{5}{|c|}{pre-test (A)}&\multicolumn{5}{|c|}{post-test (A)}\\\cline{2-21}
$\#$ &a&b&c&d&e&a&b&c&d&e&a& b& c&d&e & a& b&c&d&e \\[0.5 ex]
\hline \hline 1&13&{\it 14}&47&13&13&6&{\it 36}&39&6&13&10& {\it 33}&37&9&11&4&{\it 63}&21&6&6\\
\hline 2&27&8&33&7&{\it 25}&17&4&13&7&{\it 59}&15& 7&21&7&{\it 50}&7&2&6&4&{\it 81}\\
\hline 3&21&{\it 24}&24&19&12&5&{\it 36}&8&32&19&13& {\it 45}&16&16&10&3&{\it 61}&3&21&12\\
\hline 4&{\it 31}&14&4&23&28&{\it 42}&10&4&18&26&{\it 47}& 14&3&15&21&{\it 69}&9&3&8&11\\
\hline 5&23&41&21&{\it 4}&11&32&33&19&7&{\it 9}&15& 45&17&{\it 11}&12&20&30&12&{\it 27}&11\\
\hline 6&4&7&{\it 11}&65&13&5&6&{\it 16}&55&18&5& 4&{\it 12}&66&13&3&8&{\it 28}&37&24\\
\hline 7&21&5&19&31&{\it 24}&11&14&4&9&{\it 62}&15& 5&10&15&{\it 55}&5&4&2&4&{\it 85}\\
\hline 8&8&9&{\it 14}&48&21&12&8&{\it 13}&52&15&6& 8&{\it 30}&41&15&6&6&{\it 45}&33&10\\
\hline 9&{\it 27}&27&23&17&6&{\it 18}&34&18&9&21&{\it 26}& 33&14&12&15&{\it 36}&24&12&4&24\\
\hline 10&44&21&19&{\it 5}&11&20&24&27&{\it 15}&14&35& 20&21&{\it 13}&11&14&14&18&{\it 35}&19\\
\hline 11&16&33&19&18&{\it 14}&4&12&12&16&{\it 56}&13& 23&12&11&{\it 41}&3&6&6&6&{\it 79}\\
\hline 12&6&9&8&{\it 13}&64&7&3&13&{\it 17}&60&5& 4&9&{\it 21}&61&2&2&12&{\it 33}&51\\
\hline 13&7&32&{\it 28}&18&15&11&24&{\it 35}&15&15&6& 18&{\it 52}&10&14&7&14&{\it 53}&9&17\\
\hline 14&14&18&18&{\it 37}&13&11&14&5&{\it 52}&18&6&14&8&{\it 60}&12&5&7&3&{\it 70}&15\\
\hline 15&{\it 24}&10&55&6&5&{\it 40}&6&42&9&3&{\it 47}& 6&39&4&4&{\it 58}&2&31&5&4\\
\hline 16&20&29&{\it 29}&14&8&37&24&{\it 17}&9&13&24& 22&{\it 34}&12&8&37&19&{\it 26}&6&12\\
\hline 17&37&{\it 23}&7&26&7&28&{\it 32}&5&31&4&31& {\it 30}&5&28&6&27&{\it 46}&2&23&2\\
\hline 18&26&19&20&23&{\it 12}&12&7&23&8&{\it 50}&23& 17&19&21&{\it 20}&9&5&15&5&{\it 66}\\
\hline 19&14&{\it 22}&11&20&33&24&{\it 52}&7&10&7&14& {\it 47}&7&14&18&19&{\it 70}&3&4&4\\
\hline 20&{\it 11}&13&48&12&16&{\it 30}&12&28&15&15&{\it 35}&11&33&8&13&{\it 58}&7&12&11&12\\
\hline 21&8&23&{\it 36}&11&22&15&8&{\it 42}&19&16&13& 15&{\it 44}&11&17&10&5&{\it 62}&11&12\\
\hline 22&36&11&24&{\it 6}&23&50&10&23&{\it 7}&10&35& 7&21&{\it 16}&21&39&5&16&{\it 29}&11\\
\hline 23&32&{\it 16}&28&19&5&43&{\it 18}&29&6&4&45&{\it 21}&20&9&5&50&{\it 25}&19&4&2\\
\hline 24&{\it 6}&35&34&18&7&{\it 9}&27&37&14&13&{\it 20}&29&29&15&7&{\it 29}&24&21&20&6\\
\hline 25&20&23&18&19&{\it 20}&13&12&16&15&{\it 44}&15& 13&19&20&{\it 33}&8&9&12&12&{\it 59}\\
\hline
\hline
\end{tabular}
\caption{Distribution of alternative choices (correct choices in italics). The percent from lower
25\% (L) and percent of all students (A) who chose options (a)--(e) in the pre-test
and post-test.} \label{table2}
\end{table}

\newpage
\section*{}

\begin{figure}[h]
\includegraphics[width=3.1in]{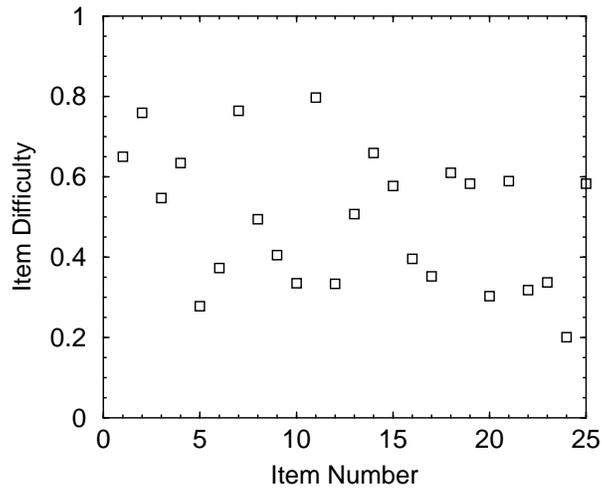}
\caption{Item difficulty (fraction correct) for each item on the test for 1170 students}
\end{figure}

\begin{figure}[h]
\includegraphics[width=3.1in]{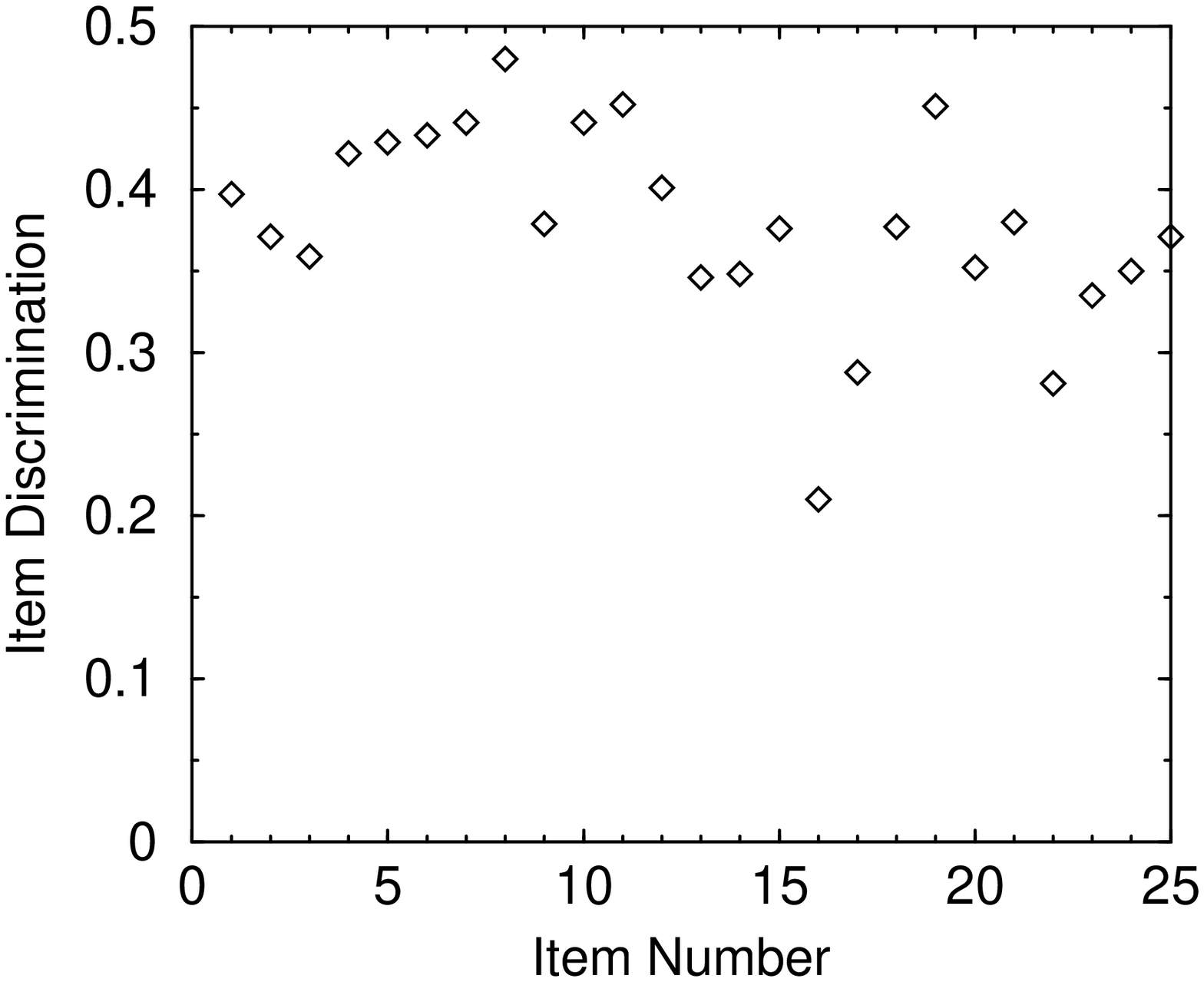}
\caption{Item discrimination for each item on the test for 1170 students}
\end{figure}


\newpage
\begin{thebibliography}{99}

\bibitem{mdt} D. Hestenes, M. Wells, and G. Swackhamer, ``Force
concept inventory,'' Phys. Teach. {\bf 30}, 141--151 (1992); R. K.
Thornton and D. R. Sokoloff, ``Assessing student learning of
Newton's laws: The force and motion conceptual evaluation and the
evaluation of active learning laboratory and lecture curricula,''
Am. J. Phys. {\bf 66}, 338--352 (1998); R. J. Beichner, ``Testing
student interpretation of kinematics graphs,'' Am. J. Phys. {\bf
62}, 750-762 (1994); R. Thornton, ``Questions on heat and
temperature,'' Unpublished, (1992); D. R. Sokoloff,
``Electric circuit concept test,'' Unpublished, (1993); D. P. Maloney, T. L. O'Kuma, C. J. Hieggelke, and A.
V. Heuvelen, ``Surveying students' conceptual knowledge of
electricity and magnetism,'' Phys. Ed. Res., Am. J. Phys. {\bf 69(7)},
S12-S23, (2001).

\bibitem{alan} A. B. Arons, {\it A Guide to Introductory Physics
Teaching} (Wiley, NY, 1990); B. A. Sherwood and W. H. Bernard,
``Work and heat transfer in the presence of sliding friction,'' Am.
J. Phys. {\bf 52}, 1001--1007 (1984); A. Van Heuvelen and X. Zou,
``Multiple representations of work-energy processes,'' Am. J. Phys.
{\bf 69(2)}, 184-194 (2001); R. A. Lawson and L. C. McDermott,
``Student understanding of the work-energy and impulse-momentum
theorems,'' Am. J. Phys. {\bf 55(9)}, 811--817 (1987); T. O'Brien
Pride, S. Vokos, and L. C. McDermott, ``The challenge of matching
learning assessments to teaching goals: An example from the
work-energy and impulse-momentum theorems,'' Am. J. Phys. {\bf 66}
(2), 147--157 (1998).

\bibitem{singh} C. Singh and D. Rosengrant, ``Students' conceptual
knowledge of energy and momentum,'' Proceedings of the Physics
Education Research conference, (Rochester, NY), edited by S. Franklin,
J. Marx, and K. Cummings, 123, 2001. 

\bibitem{nitko}G. Aubrecht and J. Aubrecht, ``Constructing
objective tests,'' Am. J. Phys. {\bf 51}, 613--620 (1983); A. J.
Nitko, {\it Educational Assessments of Students} (Prentice
Hall/Merrill, Englewood Cliffs, NJ, 1996), 2nd ed.

\bibitem{bloom} B. Bloom, {\it Taxonomy of Educational Objectives}
(Longman, New York, 1987).

\bibitem{chi} M. Chi, ``Thinking aloud,'' in {\it The
Think Aloud Method: A Practical Guide to Modeling Cognitive
Processes}, edited by M. W. Van Someren, Y. F. Barnard, and J. A.
C. Sandberg (Academic Press, London, 1994) Chap. 1.

\bibitem{factor} R. Gorsuch, {\it Factor Analysis} (Lawrence
Erlbaum, Hillsdale, NJ, 1983).

\bibitem{cohen} J. Cohen, {\sl Statistical Power Analysis for the
Behavioral Sciences} (Academic Press, NY, 1969).

\bibitem{hake} R. Hake, ``Interactive-engagement vs. traditional
methods: A six-thousand-student survey of mechanics test data for
introductory physics courses,'' Am. J. Phys. {\bf 66}, 64--74
(1998).\bibitem{chi3} M. T. H. Chi, P. J. Feltovich, and R. Glaser, ``Categorization and representation of physics knowledge by expert and novices", Cog. Sci. {\bf 5}, 121-152 (1981).\bibitem{singh2} C. Singh, ``When physical intuition fails", Am. J. Phys, {\bf 70(11)}, 1103-1109, (2002).\end{thebibliography}
\end{document}